\documentclass[aps, pre, preprint, superscriptaddress]{revtex4-1}

\draft 
\usepackage{graphicx}
\usepackage{subfig}
\graphicspath{ {./figures/} }
\usepackage{url}
\begin{document}





\title{Particles with selective wetting affect spinodal decomposition microstructures}

\author{Supriyo Ghosh}
\altaffiliation[Corresponding author. Email: supriyo.ghosh@nist.gov. Present address: ]{Materials Science and Engineering Division, National Institute of Standards and Technology, Gaithersburg, MD 20899, USA}

\author{Arnab Mukherjee}
\altaffiliation[Present address: ]{Institute of Materials and Processes, Karlsruhe University of Applied Sciences, Moltkestr. 30, 76133, Karlsruhe, Germany}

\author{T. A. Abinandanan}%
\author{Suryasarathi Bose}
\affiliation{Materials Engineering Department, Indian Institute of Science, Bangalore 560012, India}


\date{\today}
\begin{abstract}
We have used mesoscale simulations to study the effect of immobile particles on microstructure formation during spinodal decomposition in ternary mixtures such as polymer blends. Specifically, we have explored a regime of interparticle spacings (which are a few times the characteristic spinodal length scale) in which we might expect interesting new effects arising from interactions among wetting, spinodal decomposition and coarsening. In this paper, we report three new effects for systems in which the particle phase has a strong preference for being wetted by one of the components (say, A). In the presence of particles, microstructures are not bicontinuous in a symmetric mixture. An asymmetric mixture, on the other hand, first forms a non-bicontinuous microstructure which then evolves into a bicontinuous one at intermediate times. Moreover, while wetting of the particle phase by the preferred component (A) creates alternating A-rich and B-rich layers around the particles, curvature-driven coarsening leads to shrinking and disappearance of the first A-rich layer, leaving a layer of the non-preferred component in contact with the particle. At late simulation times, domains of the matrix components coarsen following the Lifshitz-Slyozov-Wagner law, $R_1(t) \sim t^{1/3}$.
\end{abstract}


\maketitle 

\section{Introduction}\label{intro}

Phase separation in ternary mixtures, which may include alloys, polymers and metallic glasses, involves a complex interplay between thermodynamics and kinetics. This complexity is further amplified by a  pre-existing ``third phase,'' which may be present in the matrix in the form of spherical particles~\cite{Lee}, a patterned  substrate~\cite{PKarim, Muthu}, a network~\cite{Chakrabarti}, a wall~\cite{Brown}, or any arbitrary shape~\cite{PWang, Sevink1999}. As particulate additives, the third phase may also be  mobile~\cite{ACBalazs, Hore, Ma, VVBalazs} or immobile~\cite{Lee, FCBalazs, DCBalazs}, and span a wide size range from tens of nanometers to microns. The presence of such a pre-existing phase introduces several new features: its interaction between the matrix components may influence the shape of the phase diagram, change the phase separation temperature and extend the miscibility window or compatibility between the phases. 

Phase separation of a binary mixture at or near a surface, referred to as surface-directed spinodal decomposition (SD), has been studied quite extensively (for a recent review see~\cite{Puri,Zeng2008,Binder2010}). The most common finding is that the surface gets enriched first with the component (say, A) with the lower surface free energy. This triggers the formation of an adjacent layer rich in B, which in turn leads to the formation of a third A-rich layer, and so on. The microstructure finally has several such alternating A-rich and B-rich layers near the surface, co-existing with the internal region with normal spinodal microstructure. This is referred as ``target pattern.'' A similar pattern is found in the case of immobile grain boundaries in polycrystalline materials~\cite{Ram2003}. However, when they are mobile, they lead to a new pattern, termed discontinuous SD~\cite{Ram2004}, in which the boundary becomes a transformation front: fast diffusion at the boundary produces alternating A-rich and B-rich lamellae perpendicular to, and behind the moving boundary, which keeps migrating into one of the (as yet untransformed) grains. 

Experimentally, a mixture can phase separate under the effect of an external field (such as shear flow~\cite{ACBalazs}, cross-linking~\cite{Qui}, electric-field~\cite{Millett2014}) or an internal field like geometrical perturbation (such as a third-phase particle~\cite{Karim}). Target patterns may be produced through either of them; for example, Tran-Cong and Harada~\cite{Qui} observed them when an external influence
(selective cross-linking reaction) triggered phase separation; Karim \emph{et al.}~\cite{Karim} observed them 
around silica nano-particles. 
Interestingly, such target pattern has also been observed in a metallic glass in which 
the C-rich particle itself was formed as a result of spinodal decomposition in a ternary blend, 
followed by a further phase separation of the C-poor (or, AB-rich) region around it \cite{Park}.

Binary mixture composition also plays a role in SD morphologies. Both experiments and simulaions have shown that a critical 50:50 mixture phase separates into bicontinuous morphologies, while droplet morphologies are prevalent in off-critical mixtures \cite{Puri,Krausch1994,PBinder,Benderly,Karim,Brown,Chakrabarti,Ma}. Two different situations were observed in off-critical mixtures based on whether the majority or the minority component is attracted to the surface. The Bulk was characterized by a bicontinuous morphology in the former and droplet in the latter. In a recent experiment by Jiang \emph{et al.}~\cite{Jiang2016}, selective interaction between the bulk components resulted in percolation which further effected the transition of bicontinuous morphology to droplet in a symmetric PMMA/SAN blend, while asymmetric blends retained their bicontinuity. The experiments by Tanaka \emph{et al.}~\cite{Tanaka} showed that when particles are mobile, phase separation could induce them to localize inside the preferred phase. The experiments by Morkved and co-workers~\cite{Mork94,Mork99} demonstrate an interesting application in which gold is made to self-assemble through selective aggregation of a phase separated mixture. In a symmetric (PS-PMMA) blend, gold particles self-assemble into PMMA phase~\cite{Mork94}, whereas in an asymmetric (PS-b-PVP) blend, gold particles self-assemble on PVP~\cite{Mork99}. The experimental work of Herzig \emph{et al.}~\cite{Herzig} and Sanz \emph{et al.}~\cite{Sanz} on the particle effects on phase separating liquid systems is particularly interesting in that it points to a novel possibility of a highly porous material produced by getting particles confined to the A-B interfaces, and draining the liquids. These studies provided interesting yet significant insights into the influence of additive surface on phase separating mixtures. However, the dependence of the morphologies on size and fraction of these particles has received little attention.

Simulations of particle effects have taken several approaches. The most common one, adopted by Lee et al~\cite{Lee},
Chakrabarti~\cite{Chakrabarti}, Millett~\cite{PWang, Millett2015}, Oono and Puri\cite{Oono1988}, and Balazs and coworkers\cite{ACBalazs,FCBalazs}, uses a Cahn-Hilliard-Cook (CHC) model for phase separation in binary blends, along with a surface interaction term at the particle-matrix interfaces. Suppa \emph{et al.}~\cite{DCBalazs} employed a lattice Boltzmann approach, in which the particles are small compared to the characteristic length scale of SD. Other approaches include Langevin particle dynamics~\cite{Mla}, fluid particle dynamics~\cite{Tanaka2006}, dissipative particle dynamics~\cite{Hore}, cell dynamics~\cite{Oono1988,VVBalazs,Ma2}, and molecular dynamics~\cite{Mac,Ma}.

In this paper, we address the role of immobile particles during spinodal decomposition with a view to elucidating interesting new microstructural features.  We use a ternary phase field model that allows us to treat the particle as a C-rich phase that co-exists with the initial binary mixture AB. The role of these particles, then, is determined primarily through how their interfaces interact with the mixture during the early stages of phase separation. This paper extends this framework in which of the role of particles on phase separation is examined through a study of the effect of the interface between the matrix and particle phases. While the matrix-particle interface could act very much like a free surface, there are two key differences: (a) the matrix-particle interface possesses a curvature, and therefore curvature-dependent phenomena such as domain coarsening (Ostwald ripening) become possible, and (b) the presence of particles at finite volume fractions introduces a new length scale: interparticle spacing, $\lambda$. In particular, when $\lambda$ is of the same order of magnitude as the spinodal wavelength, we may expect a richer variety of spinodal morphologies that arise from an interplay of phase separation, wetting and curvature-driven coarsening. 
 
In the present study, we have used a Cahn-Hilliard formulation of a ternary system. In this system, the particle phase ($\gamma$) is rich in C, with a C-poor (or, A-B rich) matrix phase separates to produce A-rich ($\alpha$) and B-rich ($\beta$) phases. The free energies of $\alpha$-$\beta$, $\beta$-$\gamma$ and $\alpha$-$\gamma$ interfaces can then be tailored easily through an appropriate choice of interaction energies, and the gradient energy coefficients. The rationale behind this work is to explore the regime where the interparticle spacing $\lambda$ is of the same order of magnitude as (but larger than) the spinodal length scale. We compare the effect of particle in two systems: one in which $\gamma$ has a strong preference for one of the phases, and the other in which they have no preference for either of the phases.

Following a description of the ternary Cahn-Hilliard model in Sec.~\ref{model}, we present our results on particle effects on SD microstructures in the neutral system and in the strongly interacting system in Sec.~\ref{results}. We discuss our results in Sec.~\ref{discussion} and summarize the main conclusions in Sec.~\ref{summary}.
\section{Model}\label{model} 
We model a binary mixture with embedded particles using a ternary system containing 
components $i$ = A, B, and C (assumed to be of similar molecular sizes). 
If local volume fractions of A ($c_A$) and B ($c_B$) are considered 
independent, then $c_C$ = $1-c_A-c_B$ becomes a dependent variable. The present work 
uses the formulation of Bhattacharyya's~\cite{Abi} ternary Cahn-Hilliard model 
where total free energy $F$ couples bulk free energy $f$ with a gradient squared term of conserved parameter $c$ as~\cite{Cahn}
\begin{equation}\label{2}
\frac{F}{K_BT}=N_V\int_{V}\left[f\left(c_A,c_B,c_C\right)+\sum_{i=A,B,C}\kappa_i\left(\nabla c_i\right)^2\right]dV, 
\end{equation}
where $N_V$ is number of molecules per unit volume and $\kappa_i$ are the bare gradient energy coefficients associated with gradients in composition of the components $i$.
Bulk or homogeneous free energy is given by regular solution expression:
\begin{equation}\label{3}
\frac{1}{K_BT}f\left(c_A,c_B,c_C\right)=\frac{1}{2}\sum_{i\neq j}\chi_{ij}c_ic_j+\sum_ic_i \ln c_i ,
\end{equation}
 where $\chi_{ij}$ is the pair-wise ($i$ and $j$) interaction parameter, $K_B$ is Boltzmann's constant and $T$ is absolute temperature. Note that $\chi_{ij}$ is inversely proportional to $T$. If a homogeneous ternary blend is quenched thermally or compositionally, it will thrust into A-rich, B-rich and C-rich domains. To track the temporal evolution of respective composition fields, the continuity equation is used:
\begin{equation}\label{4}
\frac{\partial c_i}{\partial t}= - \nabla\cdot \vec{J}_i,
\end{equation} 
$\vec{J}_i$ is net flux of component $i$. It is formulated~\cite{Ghosh} by combining results of Kramer~\cite{Kramer}, Gibbs-Duhem equations and Onsager relations. Thus we obtain the following kinetic equations for microstructural evolution: 
 \begin{eqnarray}\label{5}
\frac{\partial c_A}{\partial t}= M_{AA}\left[\nabla^2 g_A-2(\kappa_A + \kappa_C)\nabla^4 c_A - 2\kappa_C\nabla^4 c_B\right]\nonumber\\
-M_{AB}\left[\nabla^2 g_B -2(\kappa_B + \kappa_C)\nabla^4 c_B-2\kappa_C\nabla^4 c_A\right],
\end{eqnarray}
\begin{eqnarray}\label{6}
\frac{\partial c_B}{\partial t}= M_{BB}\left[\nabla^2 g_B-2(\kappa_B + \kappa_C)\nabla^4 c_B - 2\kappa_C\nabla^4 c_A\right]\nonumber\\
-M_{AB}\left[\nabla^2 g_A-2(\kappa_A + \kappa_C)\nabla^4 c_A-2\kappa_C\nabla^4 c_B\right].
\end{eqnarray} 
Here $g_A = \left(\partial f/\partial c_A\right)$ and $g_B = \left(\partial f/\partial c_B\right)$. $M_{AA}$, $M_{BB}$ and $M_{AB}$ are the effective mobilties which are given by
\begin{eqnarray}\label{7}
M_{AA}&=&\left(1-c_A\right)^2M_A+c_A^2\left(M_B+M_C\right), \nonumber\\
M_{BB}&=&\left(1-c_B\right)^2M_B+c_B^2\left(M_A+M_C\right), \nonumber\\
M_{AB}&=&\left(1-c_A\right)c_BM_A+c_AM_B\left(1-c_B\right)-c_Ac_BM_C. 
\end{eqnarray} 
Substituting $M_C$ = 0 and adjusting the matrix composition (A:B) accordingly, we obtain the effective mobilities. In all simulations, the scaled mobilities used are $M_{AA} = M_{BB} = 1.0$ and $M_{AB} = 0.98$. We start with a system containing immobile C-rich spherical particles and then allow the homogeneous matrix to phase separate following the kinetic Eqs.~\ref{5}, \ref{6}. Simulations are carried out by semi-implicit numerical integration~\cite{Shen} of the non-linear equations on a $512 \times 512$ lattice, subject to periodic boundary conditions in both $x$ and $y$ directions.
\subsection{Simulation Details}\label{details}
The particle effect on phase separation is primarily
through the strength of the interaction between the particle 
and product phases, and the interparticle
spacing $\lambda$. While we have
studied systematically the role of both these parameters, we focus
our attention on two kinds of systems: the first one, system $S_{o}$,
is neutral in terms of preference for either of the product phases 
(i.e., $\sigma_{\alpha \gamma} = \sigma_{\beta \gamma}$), and the
second, system $S_s$, in which the particle has a strong preference for the
A-rich $\alpha$ phase (i.e., 
$\sigma_{\alpha \gamma} < \sigma_{\beta  \gamma}$). We refer to this selective preference of A about the C particles as wetting. The wetting in the present scenario is solely due to relative interfacial energies between phases. In the system $S_o$, all the three interfaces have the same interfacial energy. However, in the strongly interacting system $S_s$, sum of the $\alpha$-$\gamma$ and $\alpha$-$\beta$ interfacial energies is still lower than the $\beta$-$\gamma$ interfacial energy (refer to Table 1). Thus, the $\alpha$ phase truly wets the $\beta$-$\gamma$ interface.
In our ternary phase field model, the values of the three interfacial
energies (in Tab.~\ref{ie}) are determined by the interaction parameters 
($\chi_{AB}$, $\chi_{BC}$ and $\chi_{AC}$), and gradient energy
coefficients ($\kappa_{A}$, $\kappa_{B}$ and $\kappa_{C}$) 
in Eq. \ref{2}. A short description for calculation of $\sigma$ is given in Appendix.

\begin{table}[h]
\small
\caption{Inferfacial energy of corresponding interfaces\label{ie}}
\begin{tabular*}{0.49\textwidth}{@{\extracolsep{\fill}}cccc}
\hline \hline
system & $\sigma_{\alpha\beta}$ & $\sigma_{\beta\gamma}$ & $\sigma_{\alpha\gamma}$ \\
\hline
$S_o$ & 0.15 & 0.53 & 0.53 \\
$S_s$ & 0.23 & 1.156  & 0.76  \\
\hline \hline
\end{tabular*}
\end{table}

Similarly, our results in the following section will be specifically
for two values of particle spacings: large $\lambda$ systems have
a smaller volume fraction ($V$ = 5\%) and larger particles ($R=16$), while
the low $\lambda$ systems have a higher volume fraction ($V$ = 10\%) and smaller
particles ($R=8$). These two conditions correspond to interparticle
spacings of $\lambda \simeq$ 126 and 45, respectively.

\begin{figure}[h]
\hspace{-10mm}
\begin{center}
\subfloat[]{\label{ec}\includegraphics[trim={1cm 0 5 5},clip,scale=0.3]{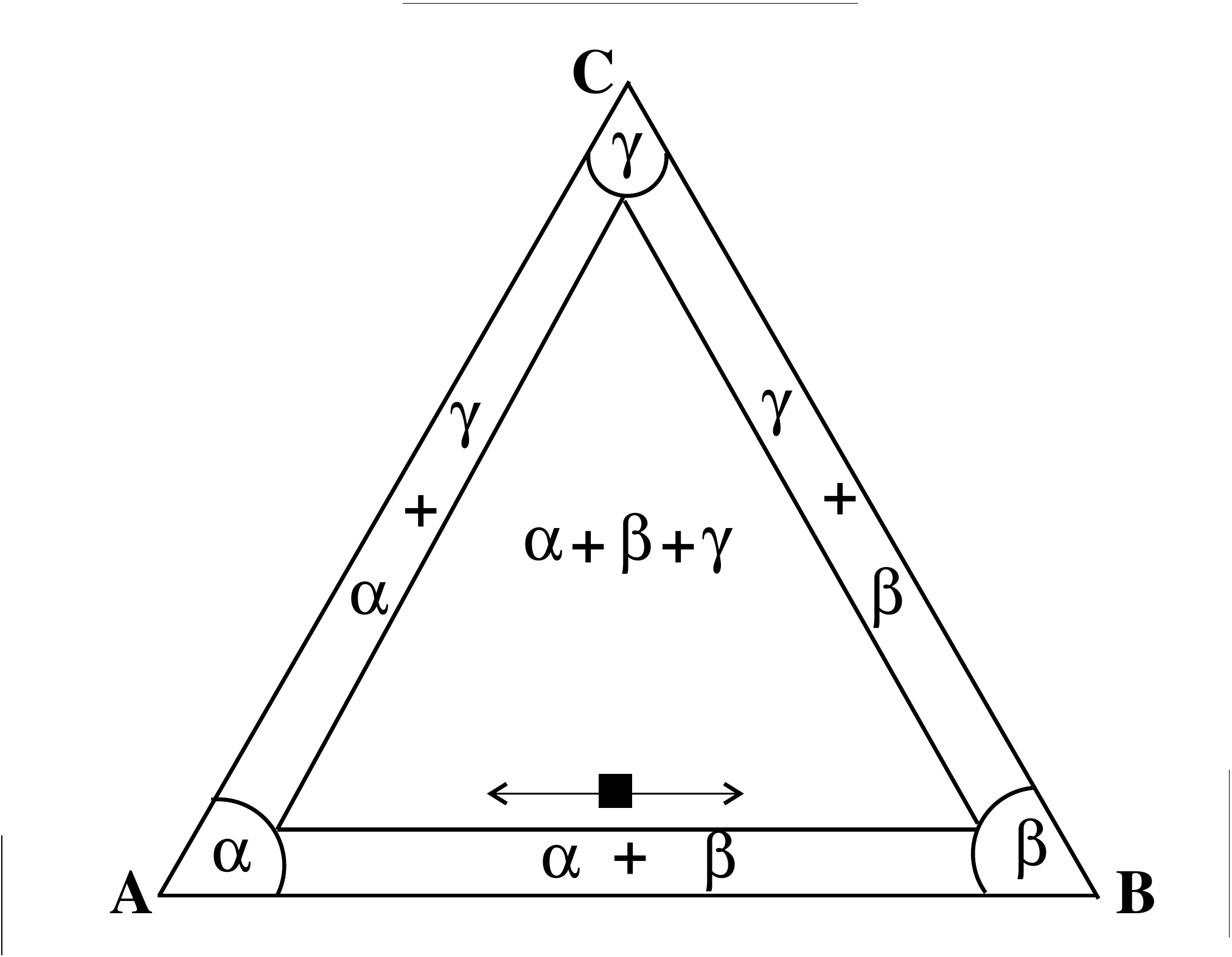}}\hspace{2mm}
\subfloat[]{\label{gs}\includegraphics[scale=0.25]{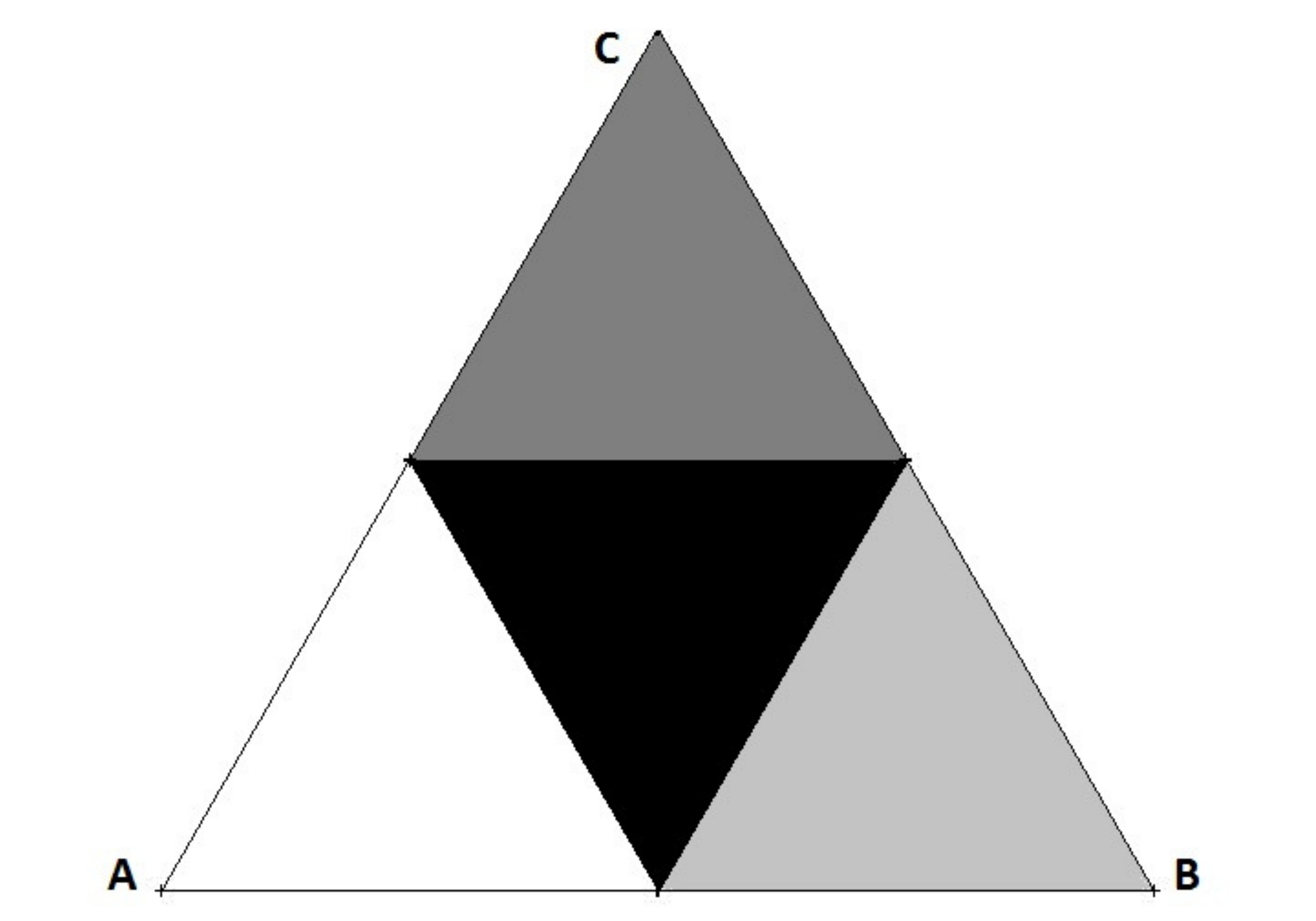}}
\caption{(a) Isothermal section of the ternary phase diagram representing $S_0$ system is depicted (schematic). The A-rich, B-rich and C-rich phases are labelled as $\alpha$, $\beta$ and $\gamma$, respectively. Simulations begin with particles having equilibrium composition of $\gamma$ ($c_A$, $c_B$, $c_C$ = 0.04, 0.04, 0.92)  and matrix having the composition in square ($c_A$, $c_B$, $c_C$ = 0.45, 0.45, 0.1). This matrix composition eventually phase separates in $\alpha$ and $\beta$ phases in the given directions (schematic).\\
(b) Gray scale color map projected on Gibbs triangle (i.e. concentration triangle). Comparing the projection of (a) on (b), C-particles are represented as dark gray and initial matrix as black which spinodally decomposes to white $\alpha$ and light gray $\beta$.
}
\label{ternary}
\end{center}
\end{figure}

In our simulations, particles start with a composition given
by that of the $\gamma$ phase in equilibrium
with $\alpha$ and $\beta$ phases in the ternary phase diagram, as shown in Fig. \ref{ec}.
Particles are also rendered immobile by making the mobility of
component C nearly zero; i.e., $M_C$ = 0 in Eq.~\ref{7}.
After randomly placing the particles in a two-dimensional 
simulation box with $512 \times 512$ grid points (with
grid spacing  $\Delta x = \Delta y = 1.0$), the matrix 
starts with a uniform composition on which a compositional noise
of $\pm 0.005$ is superimposed at each grid point. There is no noise in the particles.

Tab.~\ref{param} lists all the parameters used in our simulations.

\begin{table}[h]
\caption{Binary interaction ($\chi$) and gradient energy ($\kappa$) parameters\label{param}}
\begin{tabular*}{0.49\textwidth}{@{\extracolsep{\fill}}ccccccc}
\hline \hline
system & $\chi_{AB}$ & $\chi_{BC}$ & $\chi_{AC}$ & $\kappa_A$ & $\kappa_B$ & $\kappa_C$ \\
\hline
$S_o$ & 2.5 & 3.5 & 3.5 & 4.0 & 4.0 & 4.0 \\
$S_s$ & 2.5 & 5.0 & 3.5 & 4.0 & 8.0 & 4.0 \\
\hline \hline
\end{tabular*}
\end{table}
The local concentrations in ternary microstructures are represented using a 
gray scale map in Fig.~\ref{gs}; with this map,
$\alpha$, $\beta$ and $\gamma$ phases appear, respectively,
white, light gray and dark gray, and interfaces acquire a black edge. 

\section{Results}\label{results}

\subsection{Microstructure Evolution in System $S_o$}

\begin{figure}[h]
\centering
\subfloat[t = 3000]{\label{np_so}\includegraphics[scale=0.3]{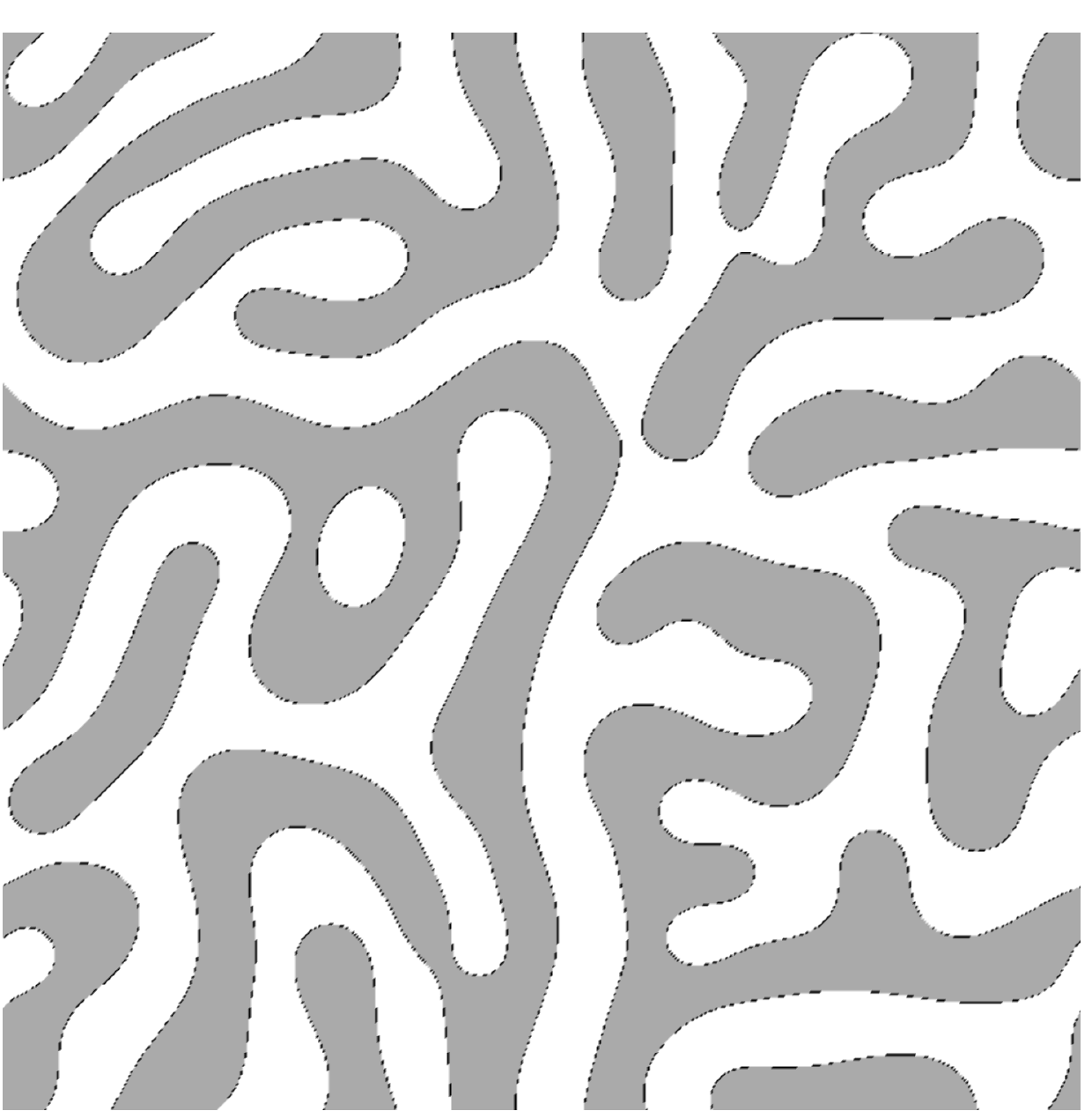}}\hspace{2mm}
\subfloat[t = 3000]{\label{sp_so}\includegraphics[scale=0.3]{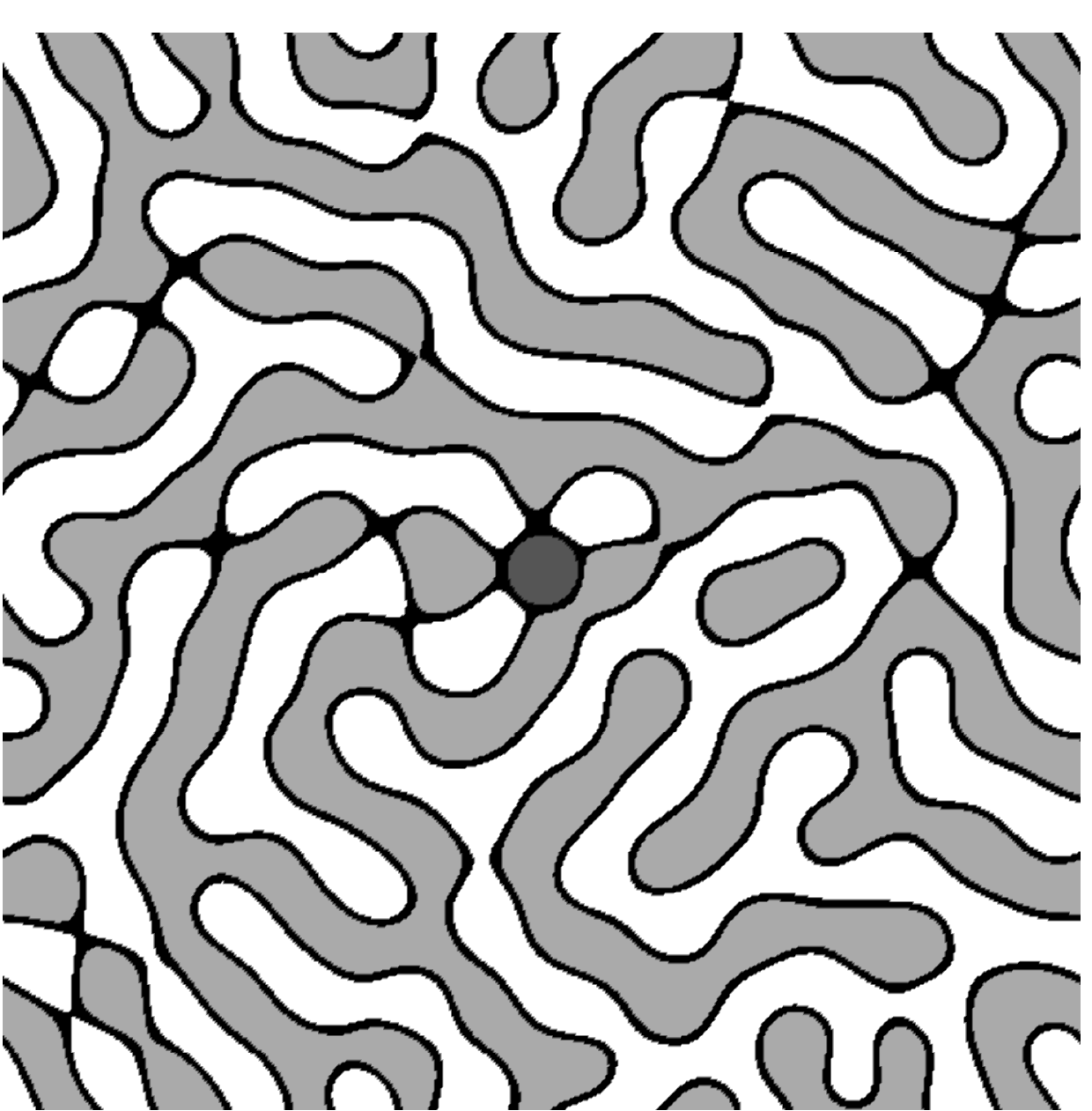}}
\caption{$A_{50}B_{50}$ : Typical microstructures in system $S_o$ (a) without particles (b) with a single particle}
\label{n1p_so}
\end{figure}

We begin with a description of spinodal decomposition in a neutral system ($S_o$) in which the particle phase has no preference for either component. In a symmetric blend in this system, phase separation leads to the well known bicontinuous microstructures shown in Fig. \ref{np_so}. In the presence of a single particle, a similar microstructure is obtained.
 
\begin{figure}[h]
\centering
\subfloat[t = 3000]{\label{so_5_16_3000}\includegraphics[scale=0.3]{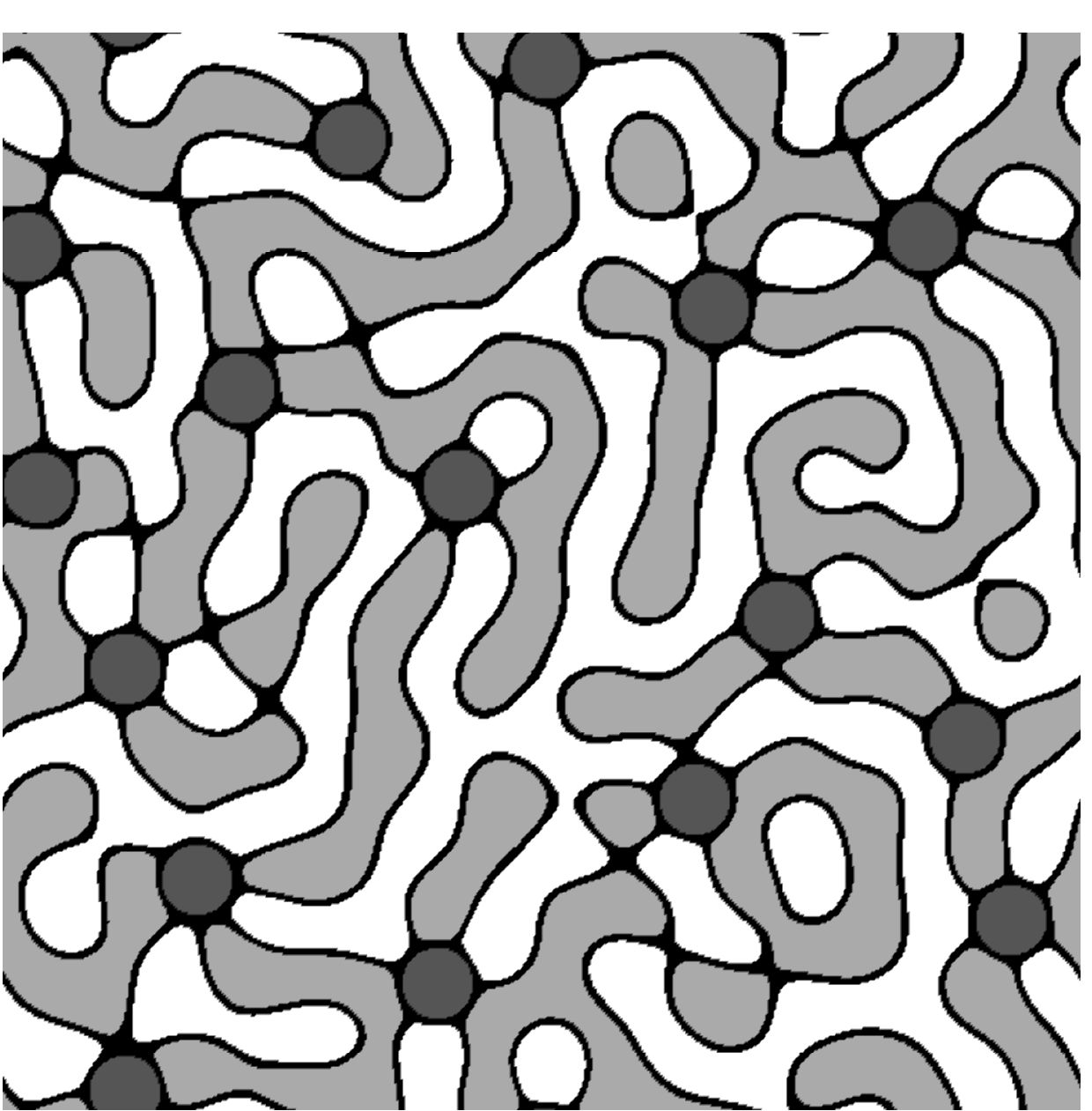}}\hspace{2mm}
\subfloat[t = 3000]{\label{so_10_8_3000}\includegraphics[scale=0.3]{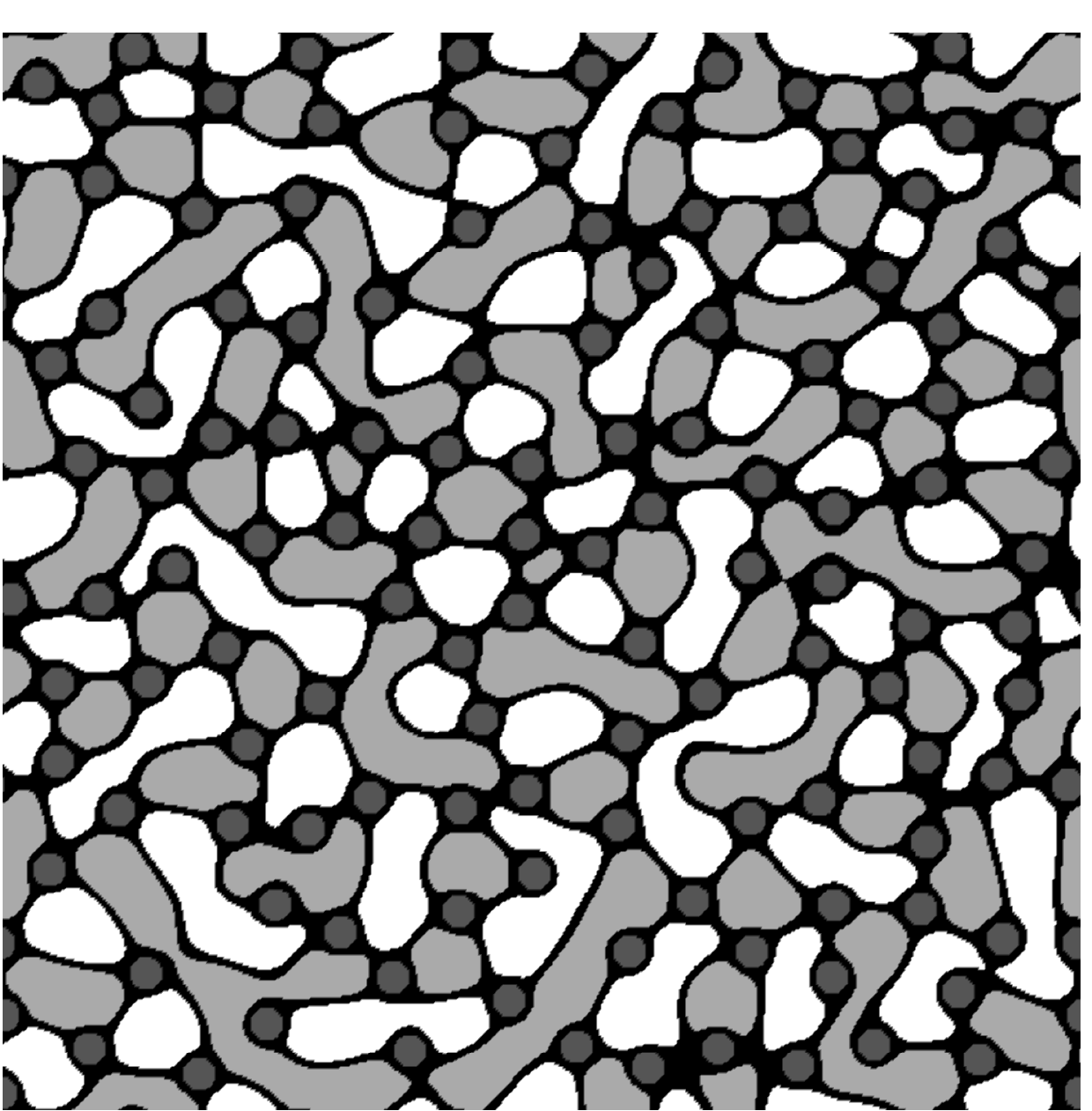}}
\caption{$A_{50}B_{50}$ : Typical microstructures for (a) $R$ = 16 and $V$ = 5\%, (b) $R$ = 8 and $V$ = 10\% using system $S_o$ parameters}
\label{so}
\end{figure}

In a system with multiple particles (at both high and low 
interparticle spacing $\lambda$), the SD microstructures 
(Fig. \ref{so}) show a nearly bicontinuous pattern.
We can easily discern the underlying bicontinuous pattern 
if we imagine replacing the particles randomly with
either of the phases in the microstructure. The main difference
between the high $\lambda$ and low $\lambda$ conditions is 
in the length scale. For example, thickness of
$\alpha$ or $\beta$ regions of the microstructure in the former
is larger than that in the latter.

\subsection{System $S_s$}

\begin{figure}[h]
\centering
\subfloat[t = 3000]{\label{np_ss}\includegraphics[scale=0.25]{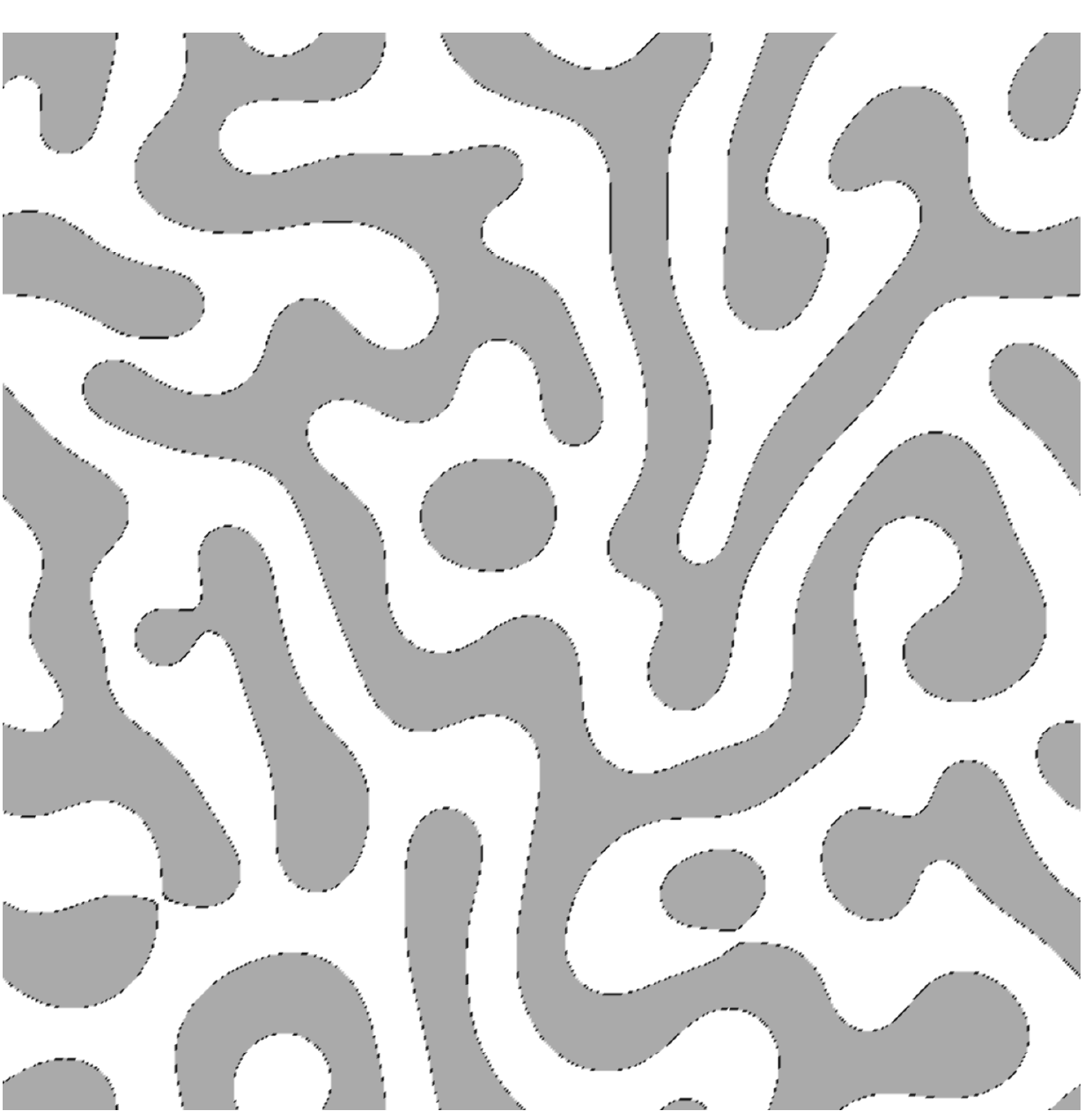}}\hspace{2mm}
\subfloat[t = 3000]{\label{sp_ss}\includegraphics[scale=0.3]{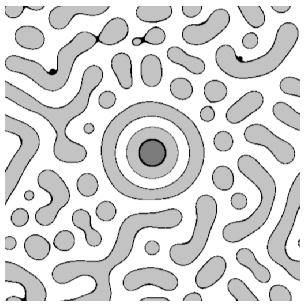}}
\caption{$A_{50}B_{50}$ : Typical microstructures in system $S_s$ (a) without particles (b) with single particle}
\label{n1p_ss}
\end{figure}

In system $S_s$, the particle phase has a strong preference for component A. This is purely due to $\sigma_{\alpha \gamma} < \sigma_{\beta  \gamma}$, which depends both on the interaction parameter $\chi$ and the gradient energy parameter $\kappa$ in Tab.~\ref{param}.
Thus, the microstructure around a single particle (in Fig.~\ref{sp_ss}) is significantly altered from that in a system with no particle (Fig. \ref{np_ss}). Specifically, we find a pattern of concentric, alternating rings of $\alpha$ and $\beta$ phases around the particle in Fig. \ref{sp_ss}. This ring pattern also referred to as a ``target pattern'', and its developments
are rationalized as follows: species of A segregates preferentially 
to the particle-matrix interface, and forms an $\alpha$ layer 
around the particle. The region around this layer gets enriched
with B, leading to the formation of a layer of $\beta$. This
process sets up a composition wave that propagates outward
from the particle~\cite{Lee}. The propagation
is arrested when the outer-most ring meets the interior that
has phase separated to a significant extent; 
therefore, the ring pattern around the particle coexists with
the normal SD microstructure in the interior~\cite{Jiang}.

As the microstructure evolves, we also find another interesting feature
in the ring pattern itself: since the rings have a curvature,
they undergo coarsening due to the Gibbs-Thomson effect~\cite{Porter}. This effect causes solute concentration adjacent to a curved surface to increase as the radius of curvature of the surface decreases. A concentration gradient therefore results, allowing the solute to diffuse in the direction of small curvature from the large, so that large curvature shrink and eventually dissolve while small curvature grow. The inner-most ring of $\alpha$ phase
has the largest curvature, and therefore, shrinks the
fastest; when this ring disappears, the particle 
finds itself surrounded by the (non-preferred) $\beta$ phase.

\begin{figure}[h]
\begin{center}
\subfloat[t = 100]{\label{5_16_100}\includegraphics[scale=0.13]{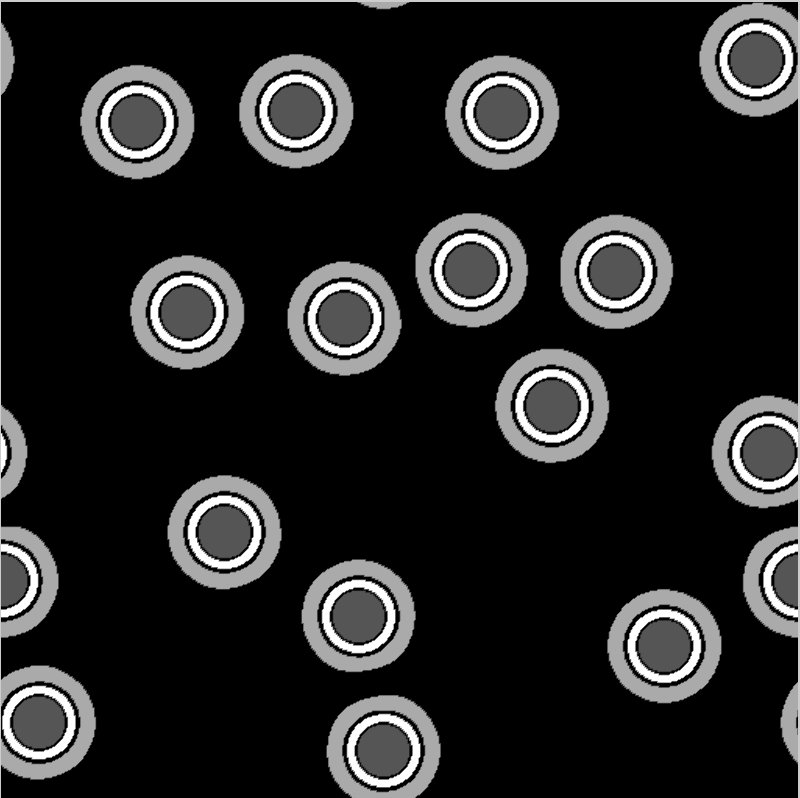}}\hspace{2mm}
\subfloat[t = 300]{\label{5_16_300}\includegraphics[scale=0.13]{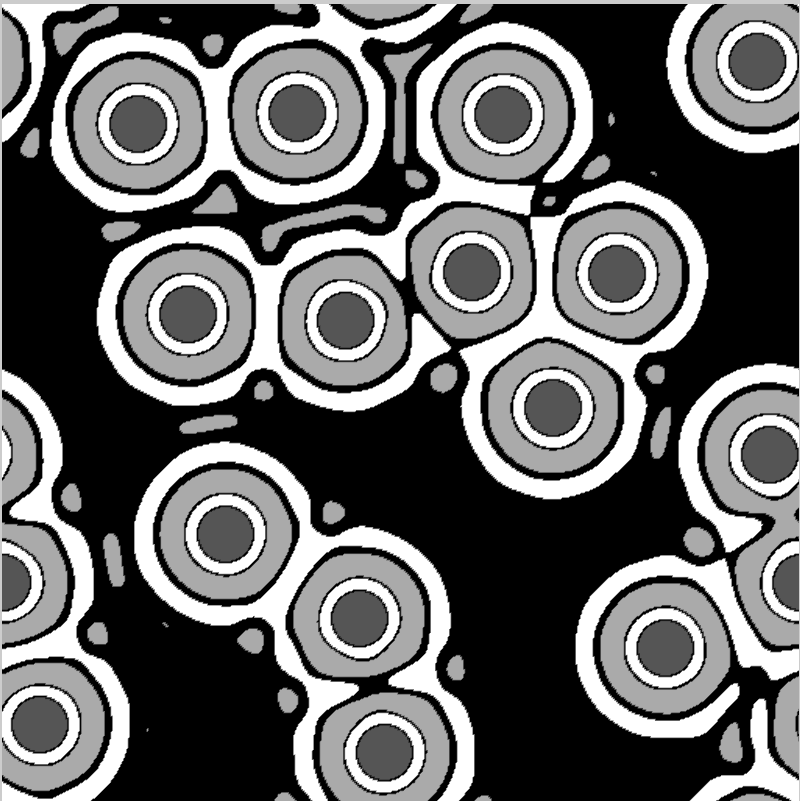}}\\
\subfloat[t = 500]{\label{5_16_500}\includegraphics[scale=0.3]{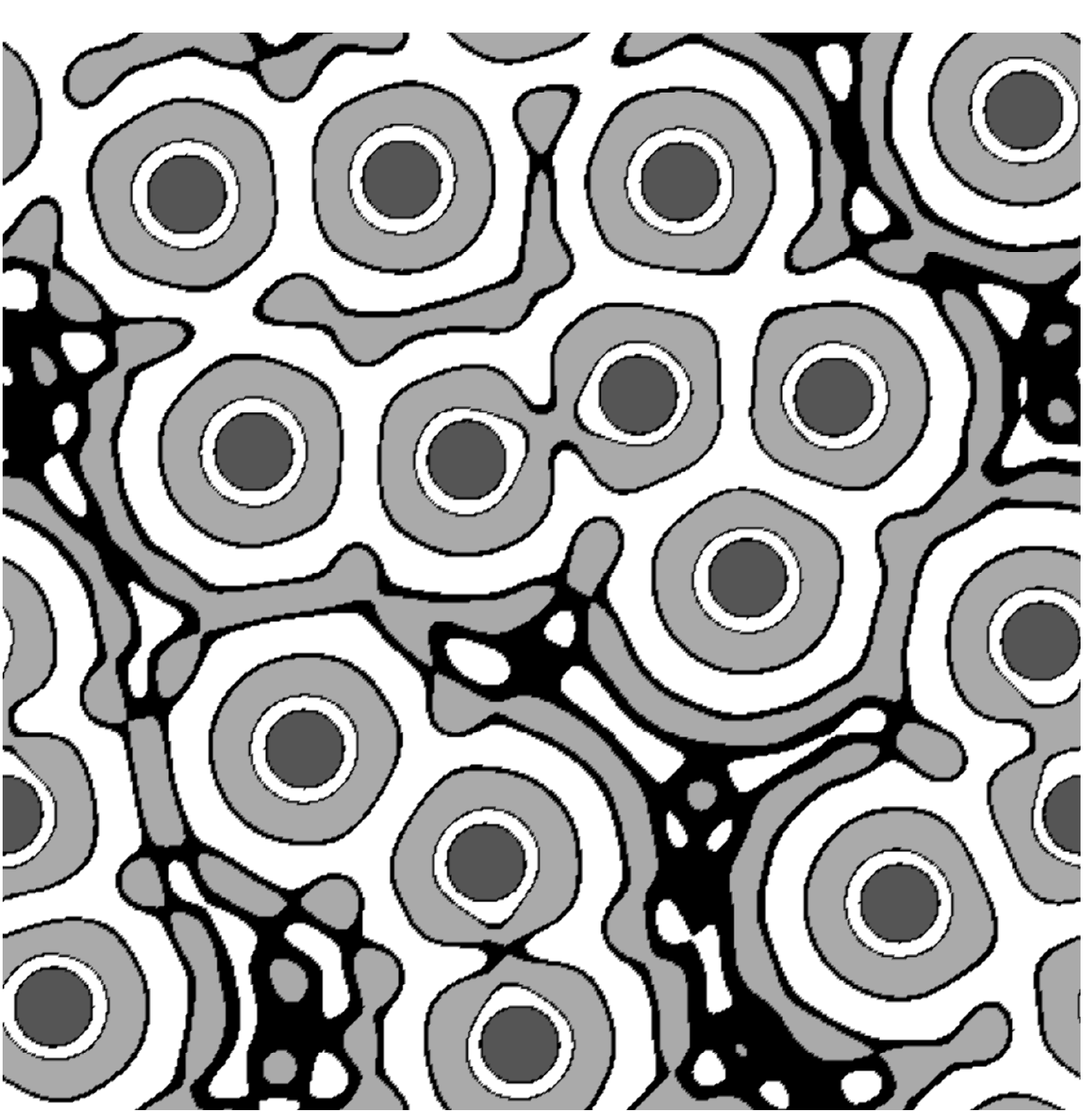}}\hspace{2mm}
\subfloat[t = 3000]{\label{5_16_3000}\includegraphics[scale=0.3]{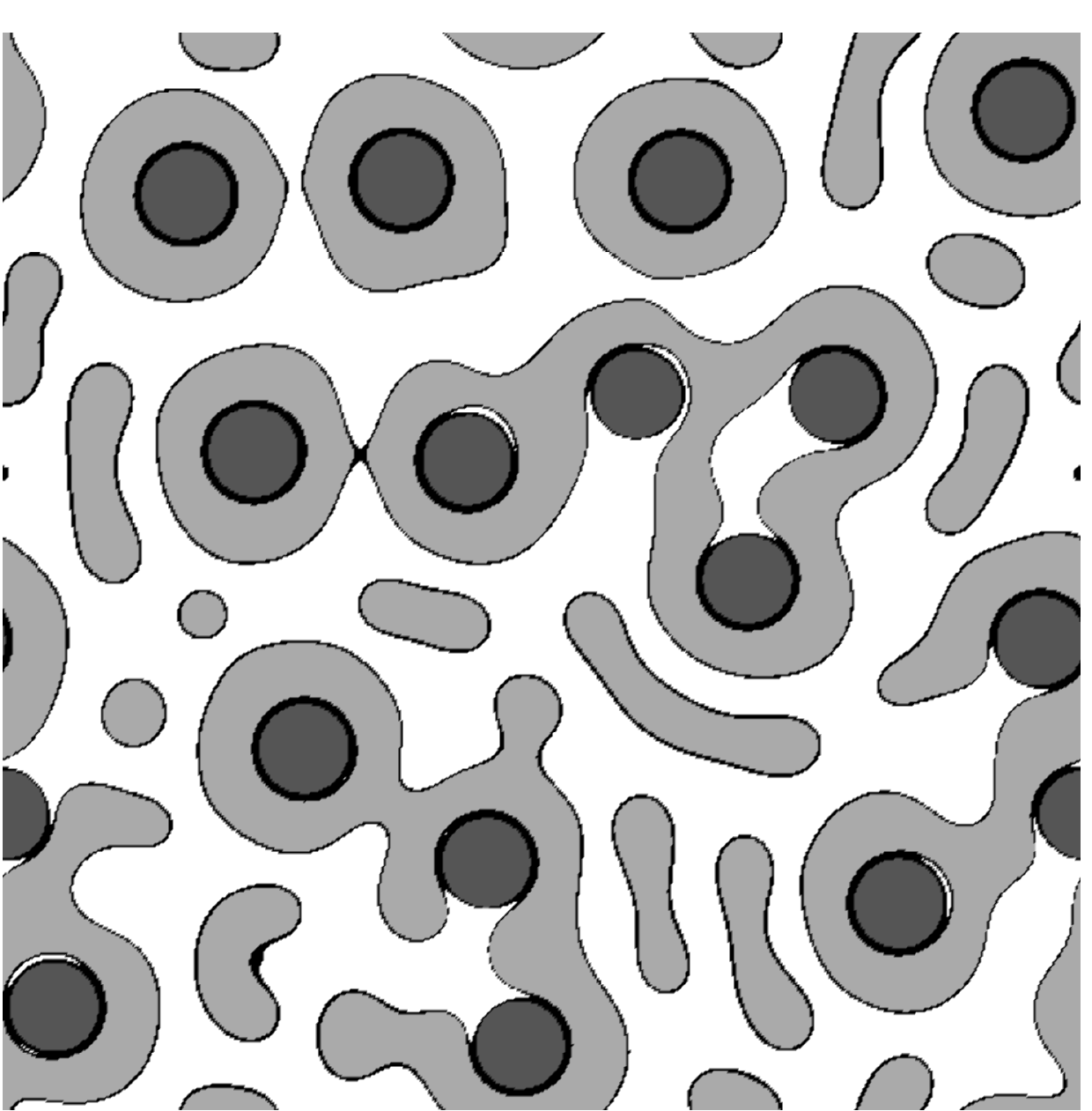}}
\caption{$A_{50}B_{50}$ : Typical microstructures from various simulation times are shown using system $S_s$ parameters. Simulations begin with particles of $R = 16$ and $V = 5\%$ distributed randomly in the homogeneous liquid. Compostion of the particles and initial liquid matrix are given in Fig.~\ref{ec}. Particles are represented as dark gray and initial matrix as black, as described in Fig.~\ref{gs}, which spinodally decomposes to white $\alpha$ and light gray $\beta$. (a, b) Concentric rings of $\alpha$ and $\beta$ initally forms around the particles. (c, d) $\alpha$ rings meet with each other on third rings, which eventually form the continuous background of $\alpha$. For the sake of brevity, we do not show the initial snapshots henceforth.}
\label{ss_c1}
\end{center}
\end{figure}

\begin{figure}[h]
\centering
\subfloat[t = 500]{\label{10_8_500}\includegraphics[scale=0.3]{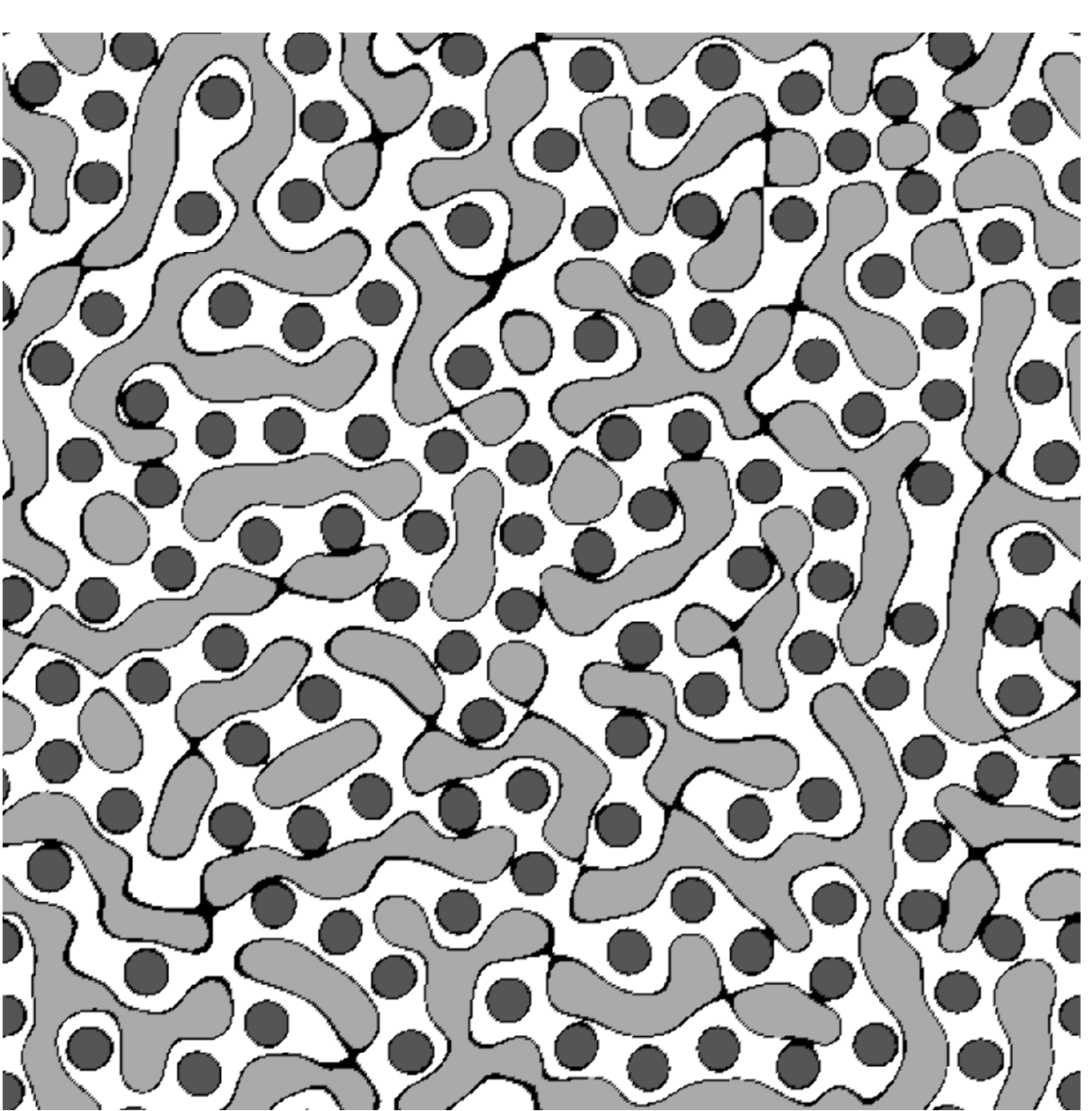}}\hspace{2mm}
\subfloat[t = 3000]{\label{10_8_3000}\includegraphics[scale=0.3]{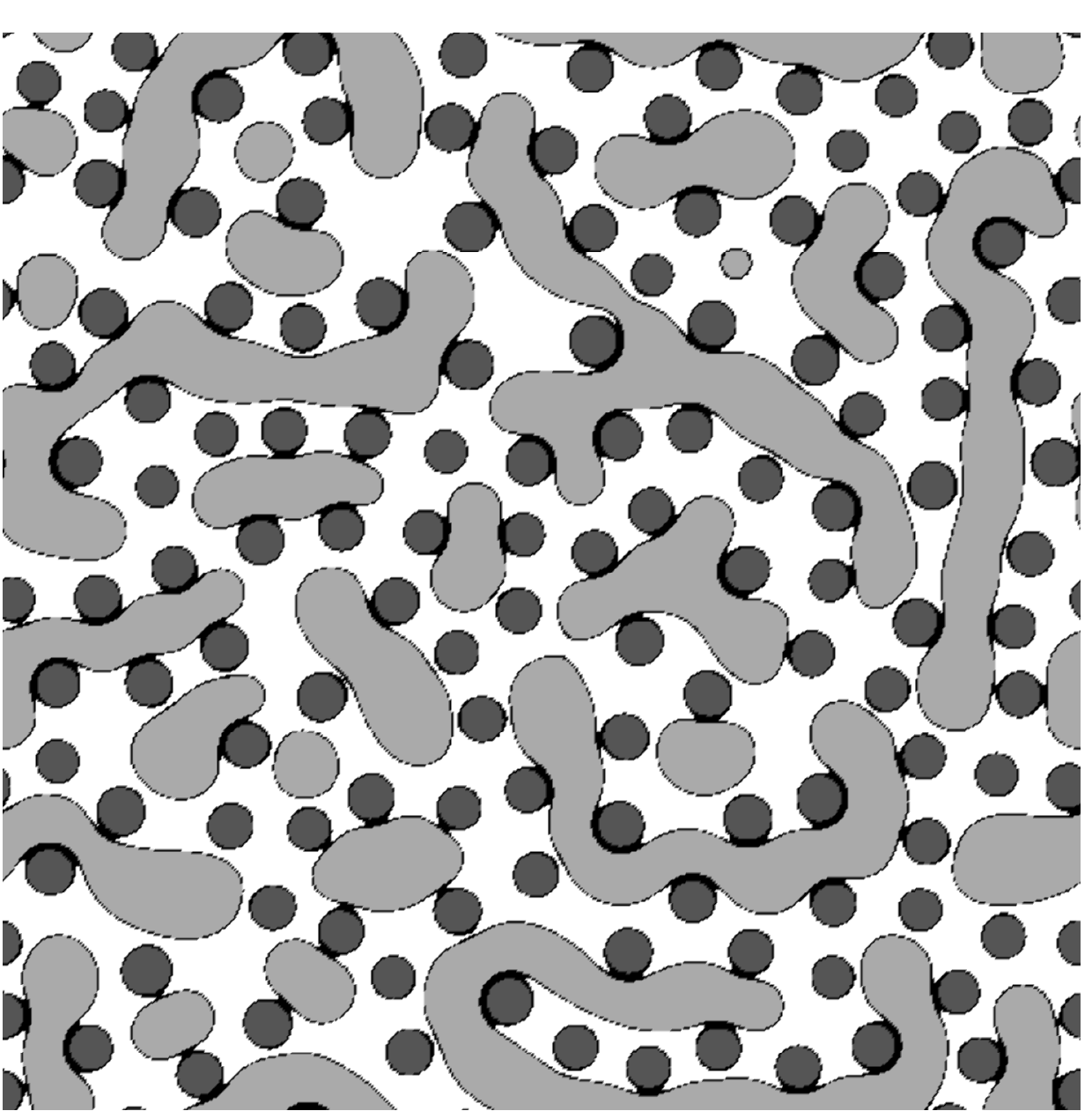}}
\caption{$A_{50}B_{50}$ : Typical microstructures for (a, b) $R$ = 16 and $V$ = 5\%, (c,d) $R$ = 8 and $V$ = 10\% using system $S_s$ parameters}
\label{ss}
\end{figure}

Surface-directed SD~\cite{Puri} would also lead to the presence of 
alternating layers of $\alpha$ and $\beta$ phases, much like
the rings in Fig. \ref{sp_ss}. However, since such layers are not
curved, the phase inversion observed in the ring microstructure is not found
in surface-directed SD.
In systems with multiple particles (at finite $V$), 
we expect the propagation of the composition wave emanating
from each particle to be stopped by those 
from neighboring particles. 
This implies that ring pattern around each particle 
would have a smaller number of rings than 
in the single particle case; this number is decided 
by the interparticle distance $\lambda$.  Thus, 
in a system with a large separation (large particles $R=16$ 
at small $V$ = 5\%), in Fig. \ref{5_16_500}), 
we find two rings around each 
particle; remnants of the third ring from each particle
have met up to form a continuous background. On the other 
hand, in the system with a small separation (see 
Fig.~\ref{10_8_500} for $R$ = 8 and $V$ = 10\%), 
the first ring of $\alpha$ phase itself comes
in contact with that from neighboring particles, with
the $\beta$ phase being confined to interparticle regions. 

Thus, at intermediate stages of phase separation, the bicontinuity is broken with the
$\beta$ islands embedded in a continuous $\alpha$ matrix. 
However, the microstructures are quite different in systems
with high $\lambda$ and low $\lambda$. 
With large interparticle separation, in Fig.~\ref{5_16_3000}, 
we find particles surrounded by just one ring of
the non-preferred $\beta$ phase, and $\beta$ islands embedded
inside the $\alpha$ matrix. With small $\lambda$, in 
Fig.~\ref{10_8_3000}, the $\alpha$ matrix has both
$\gamma$ particles and elongated $\beta$ phase islands 
embedded inside it. 

\begin{figure}[h]
\centering
\subfloat[t = 500]{\label{o1_5_16_500}\includegraphics[scale=0.3]{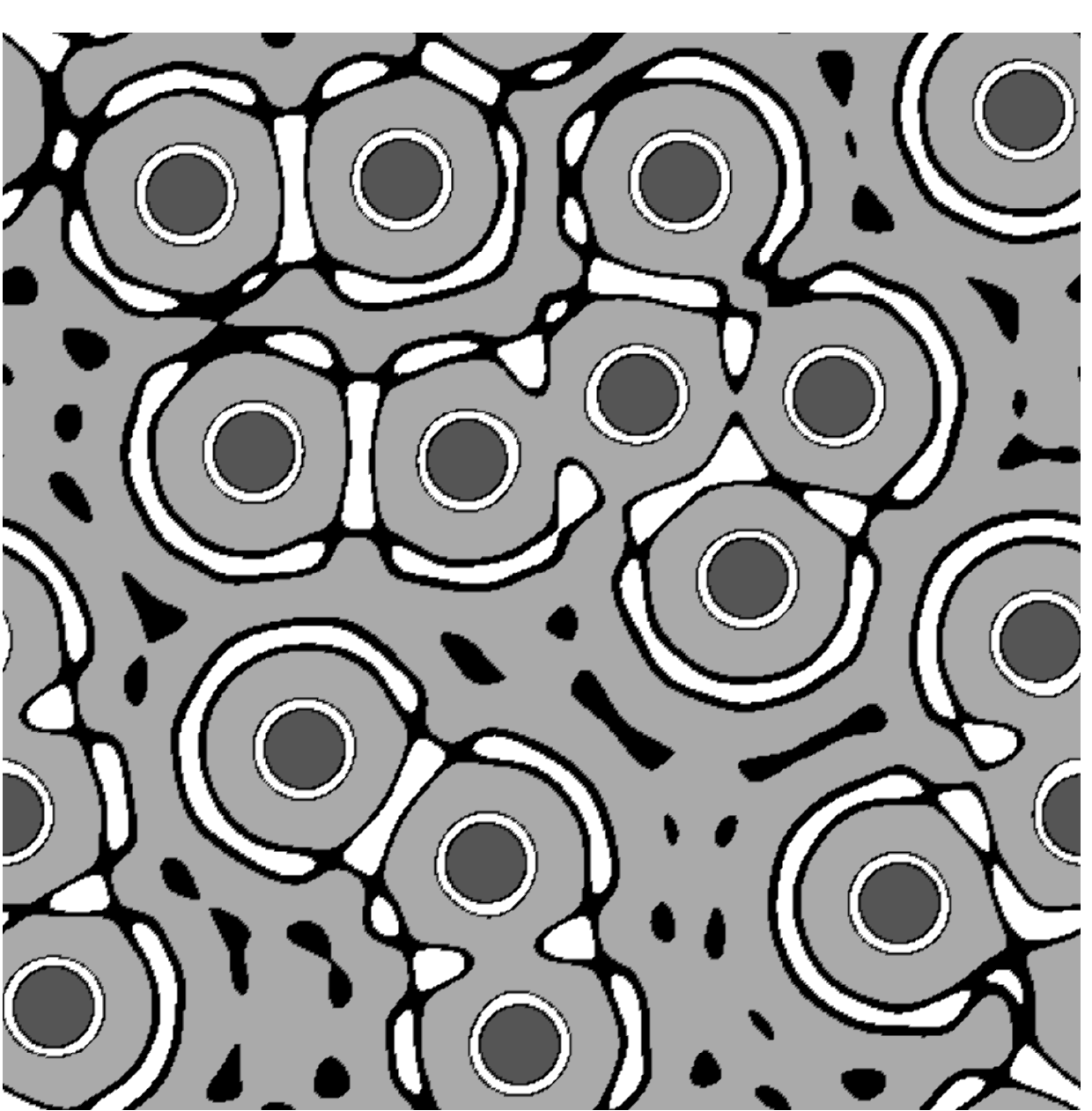}}\hspace{2mm}
\subfloat[t = 3000]{\label{o1_5_16_3000}\includegraphics[scale=0.3]{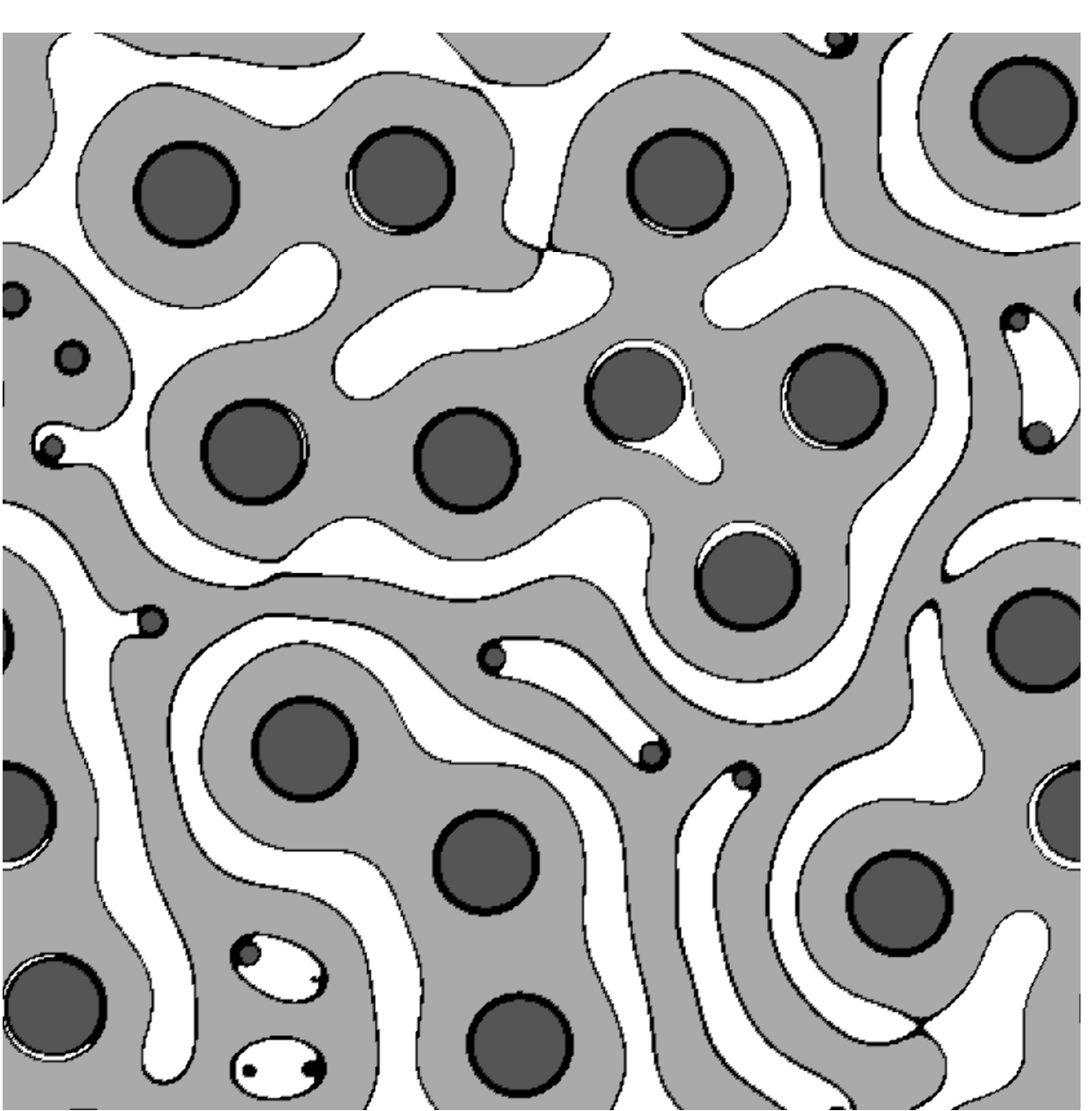}}\\
\subfloat[t = 500]{\label{o1_10_8_500}\includegraphics[scale=0.3]{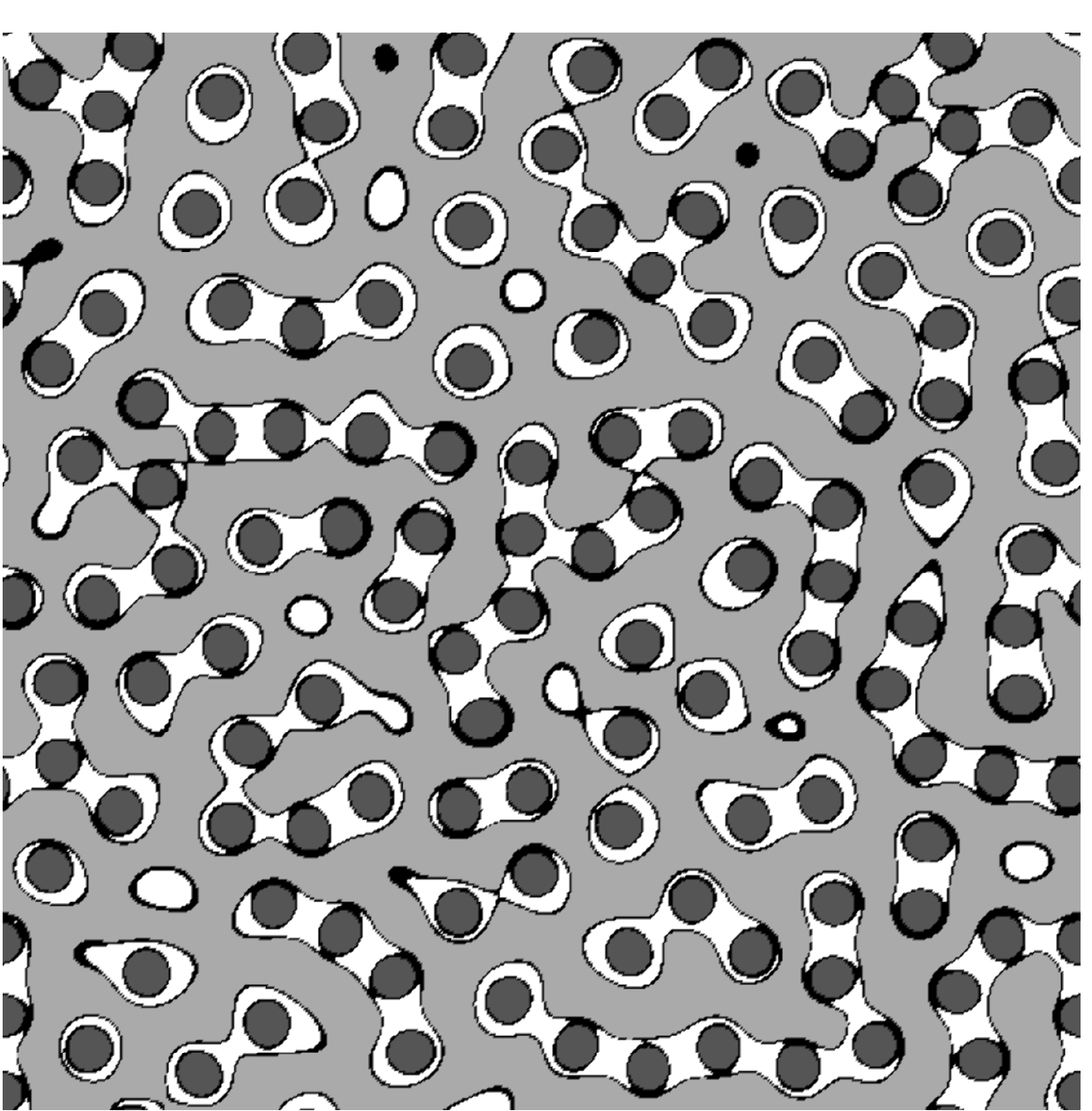}}\hspace{2mm}
\subfloat[t = 3000]{\label{o1_10_8_3000}\includegraphics[scale=0.3]{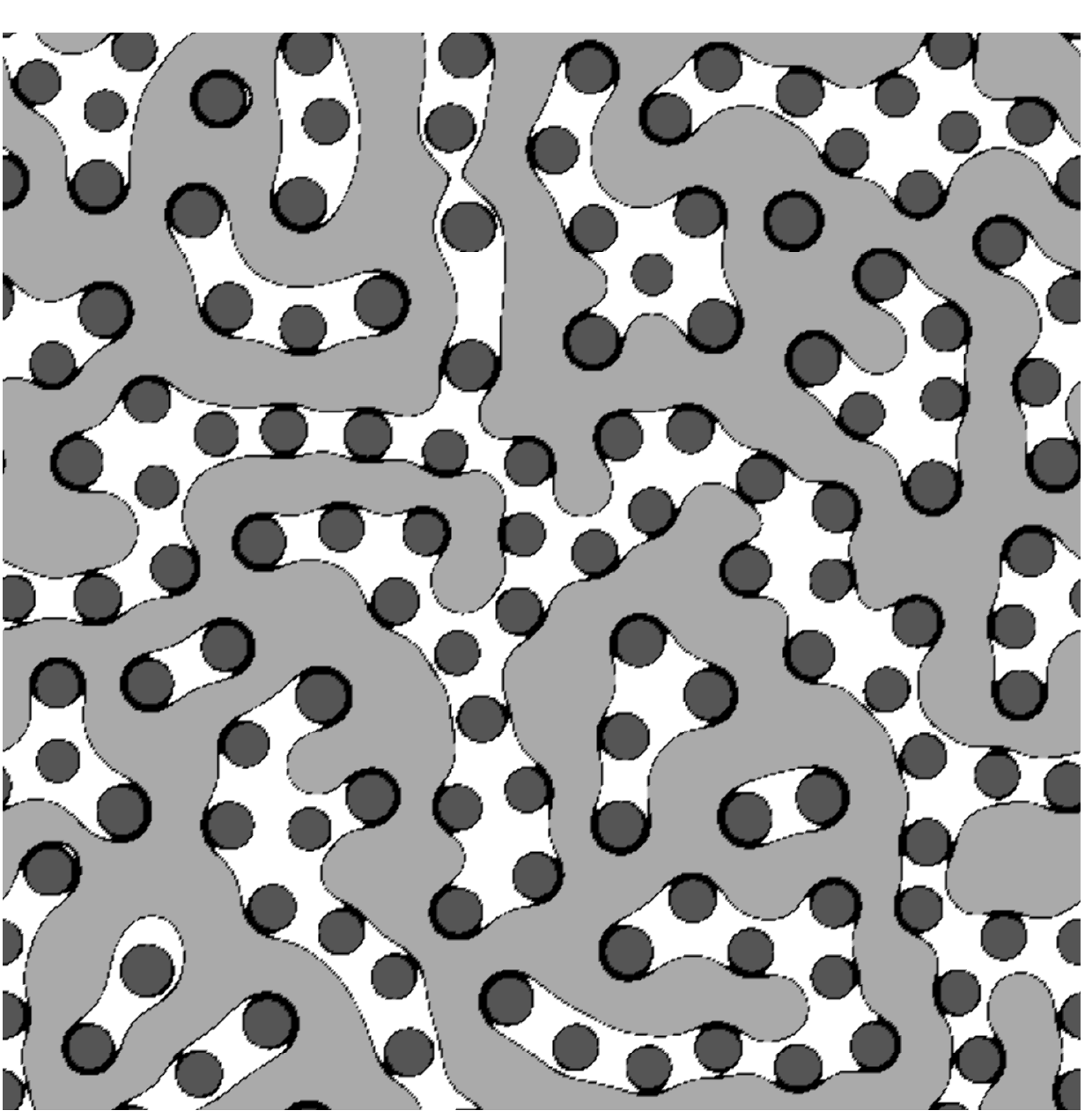}}
\caption{$A_{40}B_{60}$ : Typical microstructures for (a, b) $R$ = 16 and $V$ = 5\%, (c,d) $R$ = 8 and $V$ = 10\% using system $S_s$ parameters}
\label{o1_ss}
\end{figure}

In Fig. \ref{o1_ss}, we show the microstructural pathways in 
an asymmetric blend with $\beta$ being the majority phase.
With large interparticle spacing $\lambda$, each particle 
initially has a thin $\alpha$ and a thicker $\beta$ ring around it, 
with the remaining $\alpha$ forming a thin, meandering, 
river-like feature in the interparticle regions. When $\lambda$ is
small, however, the ring patterns around particles impinge 
at the first or the second ring; 
when they impinge at the first ring (e.g., around 
closely spaced particles), we find chains of $\gamma$ particles
inside the $\alpha$ phase. At other places, we find 
the $\beta$ as a continuous phase. What is noteworthy, however, 
is that, at intermediate times, the $\alpha$ phase regions (with
the particles inside them) become interconnected (e.g., by
piercing the continuous $\beta$ phase at its thinnest parts),
and develop bicontinuity of $\beta$ and $\alpha$ phases in Fig. \ref{o1_10_8_3000}.

\begin{figure}[h]
\centering
\subfloat[t = 500]{\label{o2_5_16_500}\includegraphics[scale=0.3]{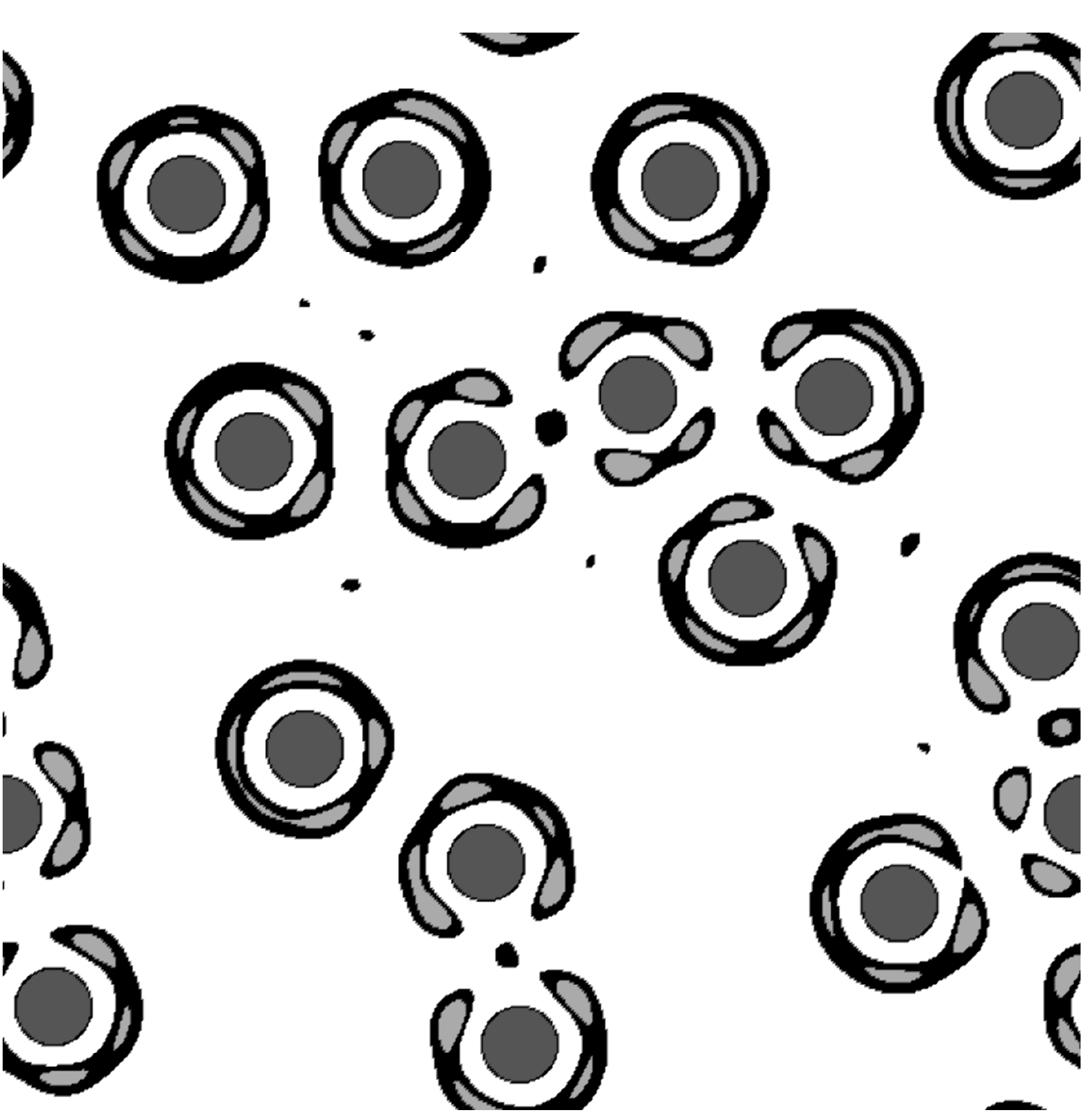}}\hspace{2mm}
\subfloat[t = 3000]{\label{o2_5_16_3000}\includegraphics[scale=0.3]{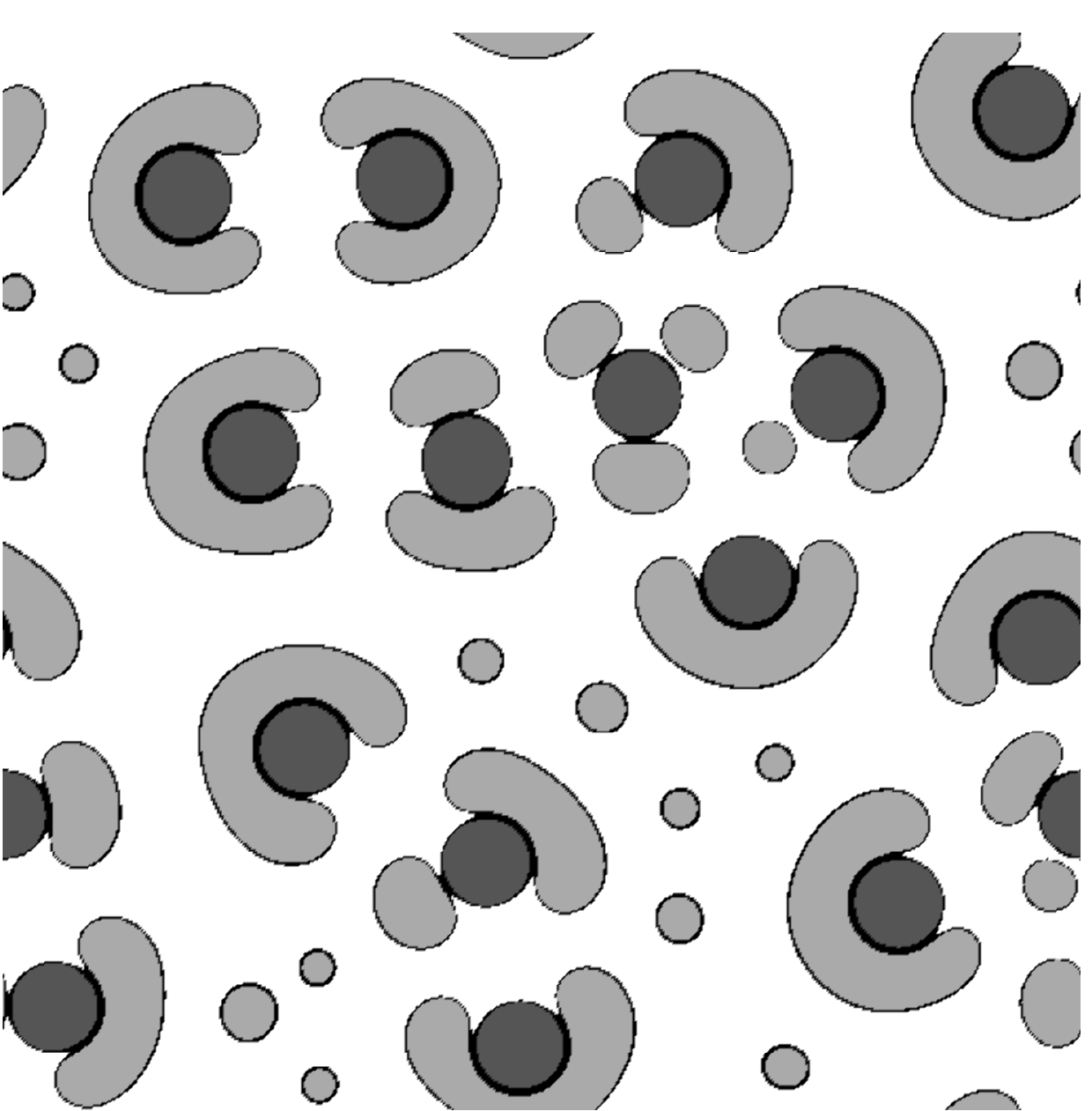}}\\
\subfloat[t = 500]{\label{o2_10_8_500}\includegraphics[scale=0.3]{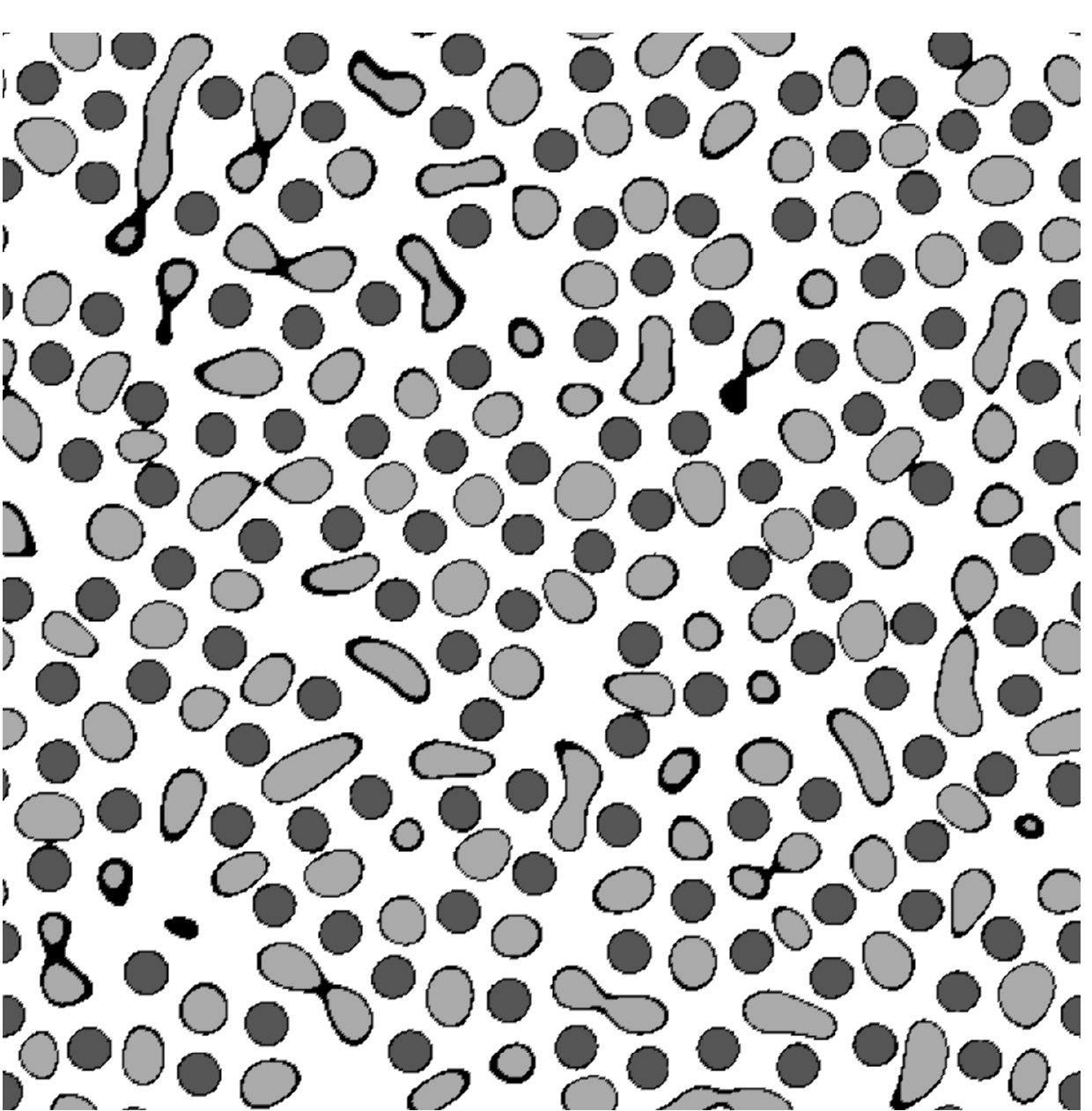}}\hspace{2mm}
\subfloat[t = 3000]{\label{o2_10_8_3000}\includegraphics[scale=0.3]{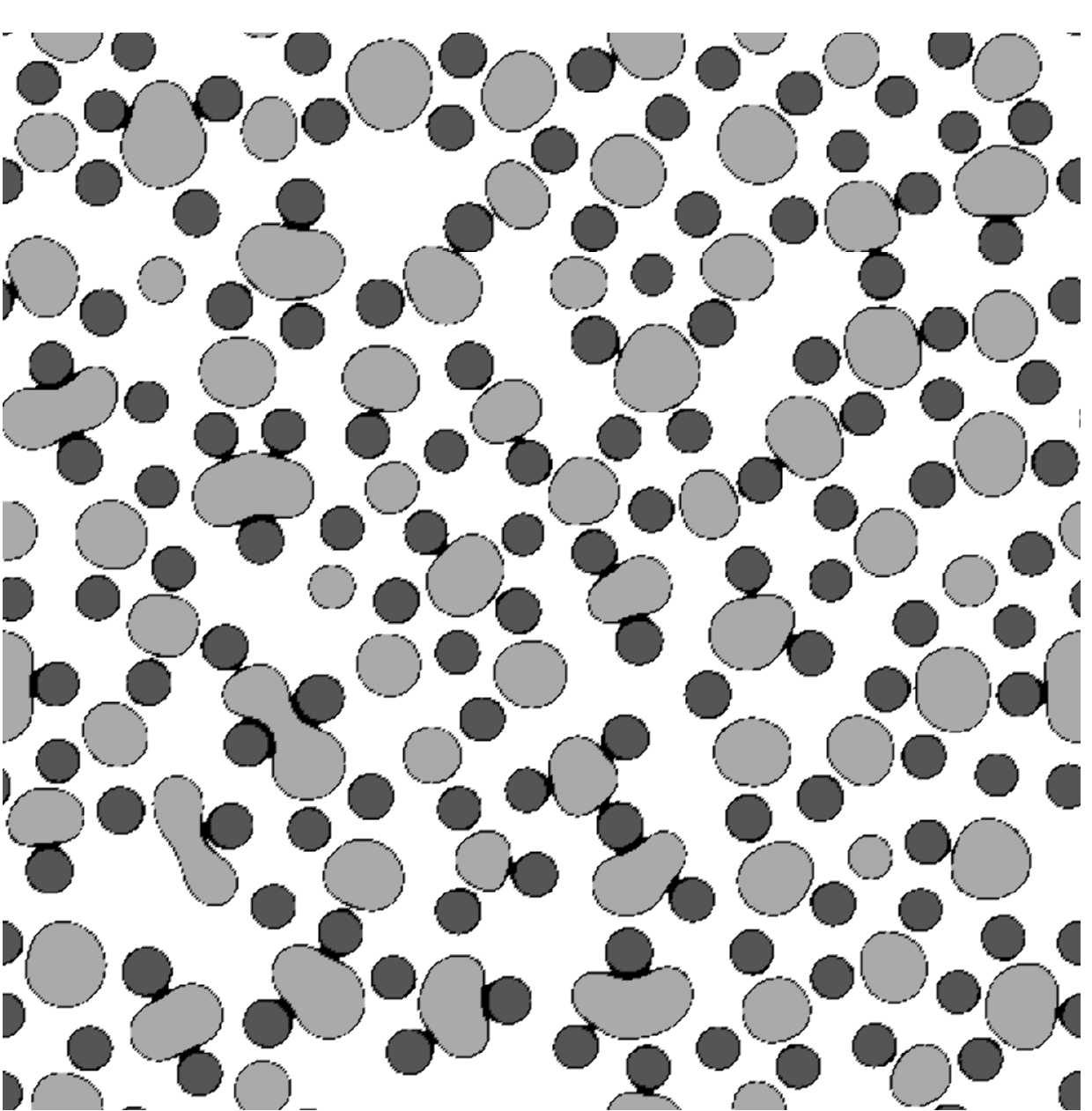}}
\caption{$A_{60}B_{40}$ : Typical microstructures for (a, b) $R$ = 16 and $V$ = 5\%, (c,d) $R$ = 8 and $V$ = 10\% using system $S_s$ parameters}
\label{o2_ss}
\end{figure}

For the sake of completeness, we present the microstructural
pathways in blends with $\alpha$ phase as the majority phase (Fig. \ref{o2_ss}),
which is also preferred at the particle matrix interface in
system $S_{s}$. The microstructural
development promotes the formation of islands of the minority
$\beta$ phase embedded in a continuous $\alpha$ phase. At 
$t=3000$, the microstructures in the large $\lambda$ system
exhibits $\beta$ phase islands surrounding the particles, 
clearly as a result of their origin as a second ring. In the 
small $\lambda$ system, however, small $\beta$ islands are 
formed right at the beginning at the interparticle regions.

\subsection{Coarsening kinetics of the wetting phase $\alpha$}
Domain growth in the above microstructures is characterized by a time-dependent structure function, $S(\textbf{k},t)$ \cite{chakrabarti1989late, zhu1999coarsening}. In case of ternary systems, there exist three linearly independent structure functions \cite{hoyt1989spinodal, hoyt1990linear} and assuming evolution process to be isotropic the circularly averaged structure factor in the $xy$-plane with $N$ lattice points is given by
\begin{equation}
S_{ii}(k,t) = \frac{1}{N}\left\langle c_{i}^*(\textbf{k},t)c_{i}(\textbf{k},t)\right\rangle,
\end{equation}
where $k$ is the magnitude of the wave vector $\textbf{k}$. The $k$ value corresponding to the maximum of $S_{ii}(k,t)$ is a measure of domain size. Precise location of this maximum however is difficult to extract due to the discretization involved in the phase-field simulations. Domain size is therefore monitored through some moment of the structure function, usually the first or the second. Here we use the first moment of $S_{ii}$ to represent the average size of the A-rich $\alpha$ domains (or B-rich $\beta$ domains) which is given by
\begin{equation}\label{eq_first_moment}
R_1(t) = \frac{1}{k_{1}(t)} = \frac{\sum S_{ii}(k,t)}{\sum k S_{ii}(k,t)}.
\end{equation}
At intermediate to late stages of spinodal decomposition, bulk domains supposed to grow following the Lifshitz-Slyozov-Wagner (LSW) law, which gives $R_1(t) \sim t^{1/3}$ in case of binary mixtures due to the diffusion~\cite{lifshitz1961kinetics, Wagner1961}. 

\begin{figure}[h]
\centering
\subfloat[]{\includegraphics[scale=0.5]{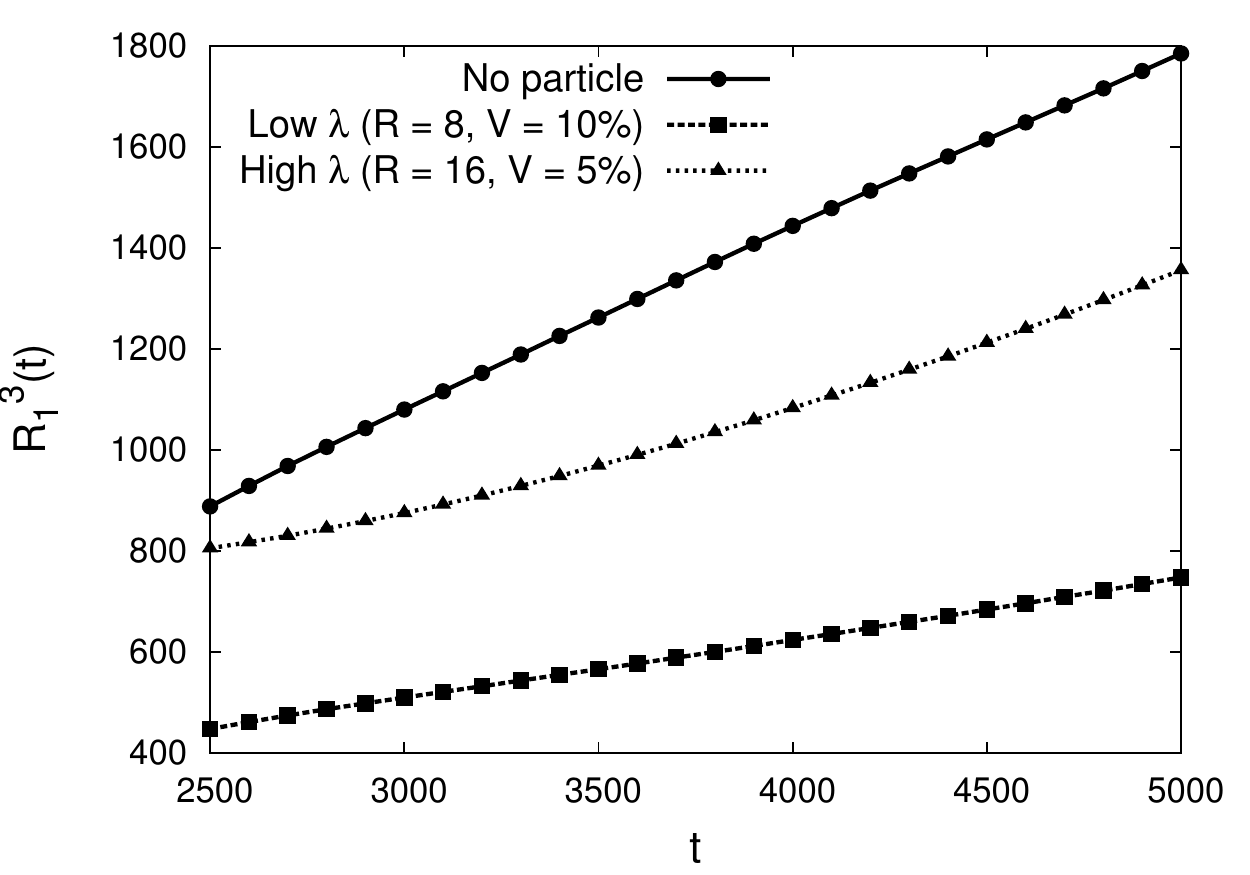}} \hspace{2mm}
\subfloat[]{\includegraphics[scale=0.5]{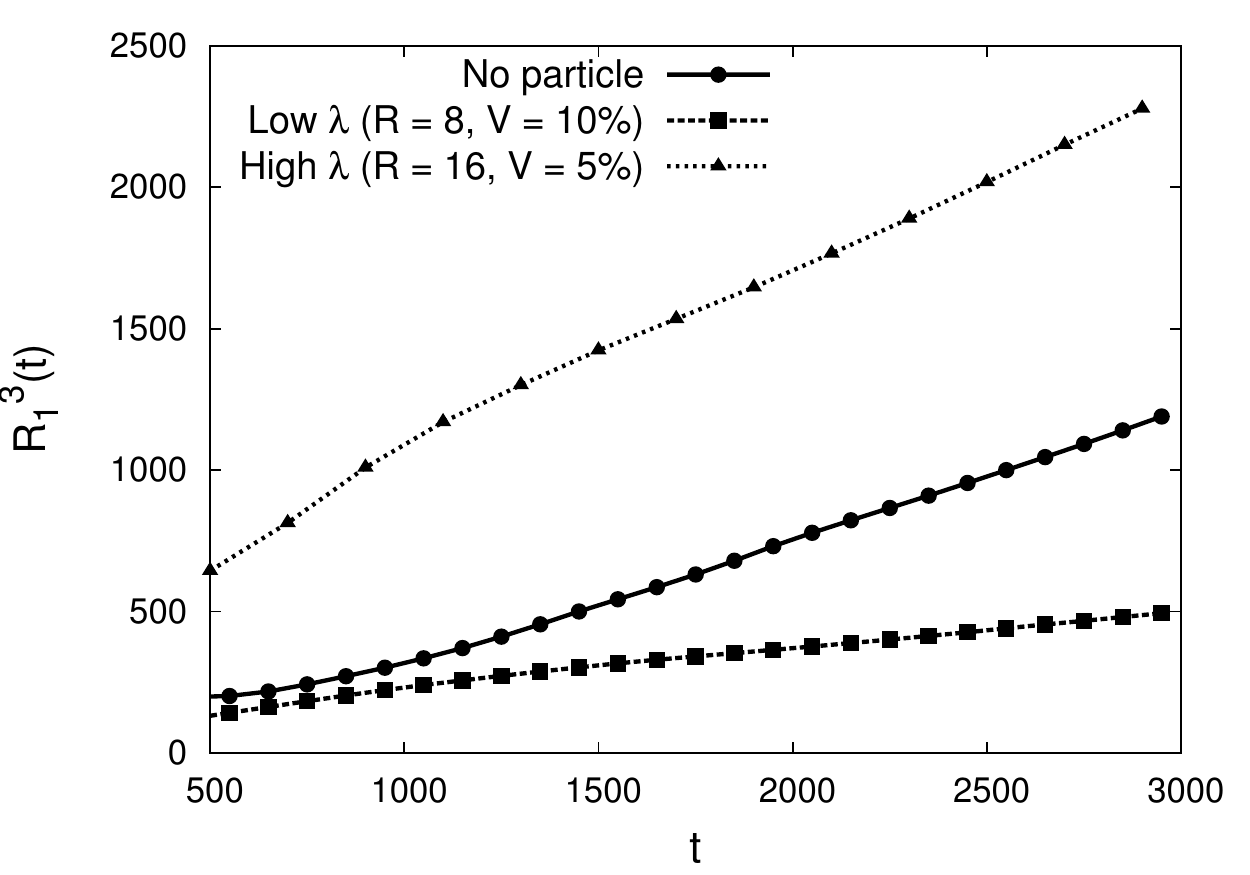}}
\caption{\label{5050}$A_{50}B_{50}$: Average size of the $\alpha$ domains $R_1(t)$, given by Eq.~\ref{eq_first_moment}, is plotted against time $t$ in (a) $S_o$ and (b) $S_s$ systems. Note that domain growth follows the LSW $t^{1/3}$ law. Note that timescales are different in the above figures.}
\end{figure}

In the above reference, $R_1(t)$ of the $\alpha$ domains is presented in Fig.~\ref{5050} for the symmetric case. Several interesting points can be deduced from this. First, in the absence of particles, coarsening kinetics are consistent with the LSW law. Though the domain size as well as the coarsening rates are markedly affected by the presence of particles indicated by different slopes of the curves, the coarsening law in itself is not significantly altered in both $S_o$ and $S_s$ systems.

Second, in the $S_o$ system, presence of particles suppress the coarsening rates, the effect being more prominent with low $\lambda$, leading to a smaller $R_1(t)$. Spinodal decomposition proceeds via amplification of the composition fluctuations with maximal growing wavelength given by $\lambda_{SD} = \sqrt{\frac{16\pi^2(\kappa_{i}+\kappa_{j})}{-\partial^2 f/\partial c^2}}$, after Cahn~\cite{cahn1961spinodal}. Since the spinodally decomposed matrix is essentially binary $\alpha\beta$, the above expression yields $\lambda_{SD} \approx 36$. Note that $\lambda_{SD}$ is comparable to the small $\lambda$ used in the phase-field simulations, whereas large $\lambda$ used is $\approx 3.5$ times $\lambda_{SD}$. As the bulk domains grow beyond $\lambda_{SD}$, the subsequent coarsening is dictated in the scale of $\lambda$, which is represented by the density of particles and thus acts as a constraint. While growth is hindered from early stages with low $\lambda$, the domains are able to coarsen at a rapid rate with high $\lambda$, before the particles ``see'' the phase separation.

\begin{figure}[h]
\centering
\subfloat[]{\includegraphics[scale=0.5]{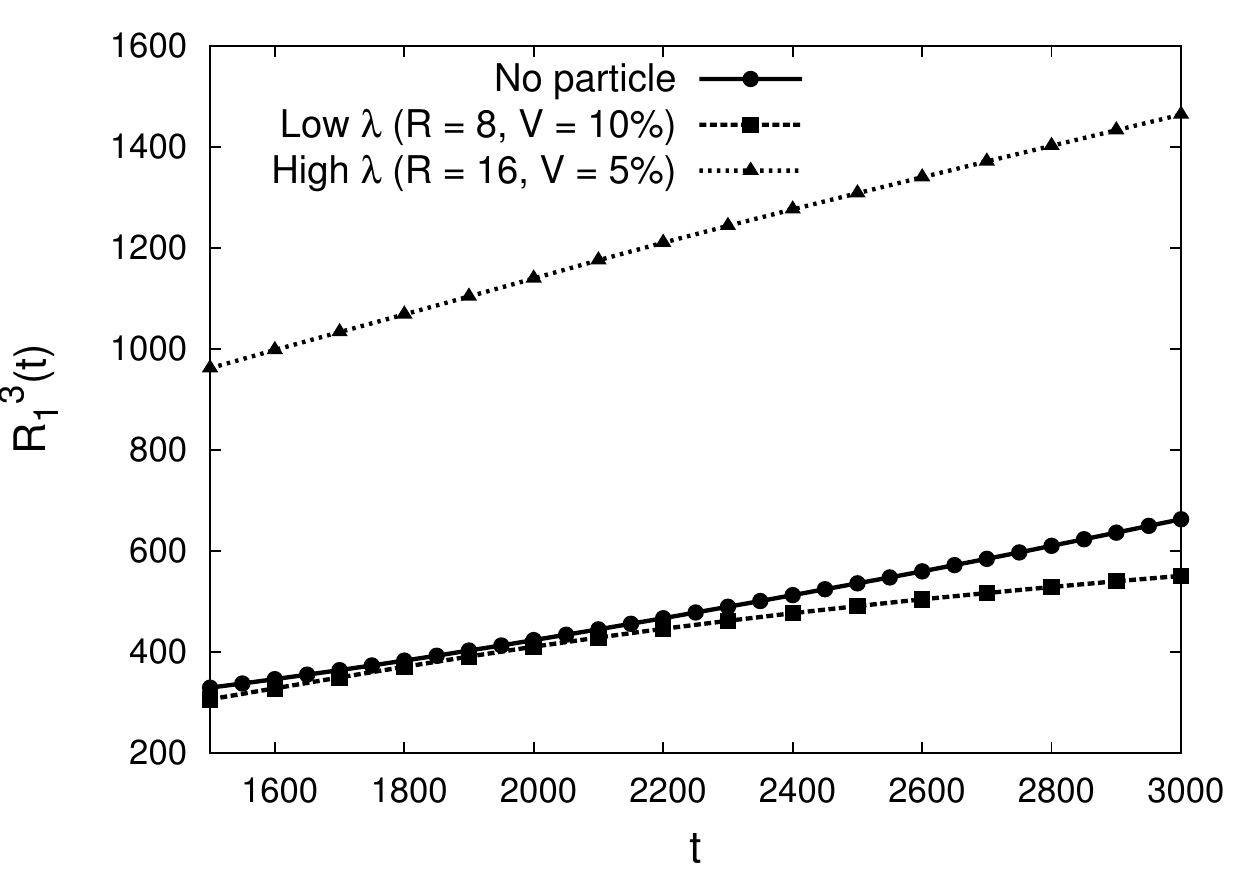}}\hspace{2mm}
\subfloat[]{\includegraphics[scale=0.5]{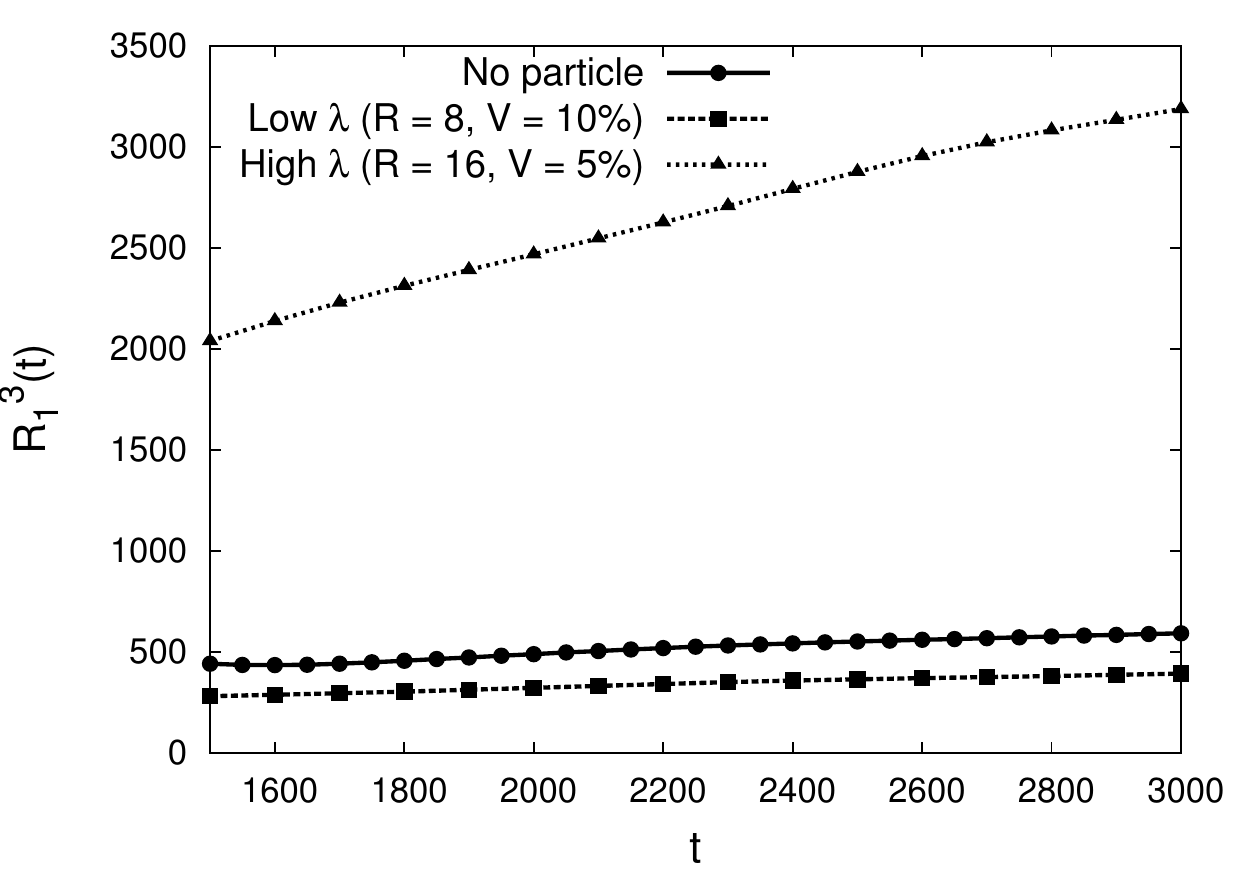}}
\caption{\label{4060} Average size of the $\alpha$ domains $R_1(t)$, given by Eq.~\ref{eq_first_moment}, is plotted against time $t$ in (a) $A_{40}B_{60}$ and (b) $A_{60}B_{40}$ $S_s$ systems. Note that domain growth follows the LSW $t^{1/3}$ law.}
\end{figure}

Third, the above arguments do not hold in the $S_s$ system, where, interestingly, coarsening in the large $\lambda$ system overwhelms that of neat blends. This indicates that in addition to $\lambda$, wetting also plays a significant role in the kinetics of coarsening. Wetting-induced phase separation proceeds via formation of concentric rings of preferred $\alpha$ and non-preferred $\beta$ phases around the $\gamma$ particles. The length scale in such a network is governed by the rapid growth of the wetting rings and then merging of these rings about the adjacent particles which subsequently extends into the background spinodal pattern. Referring to Figs.~\ref{5_16_500} and \ref{10_8_500}, with high $\lambda$ the third rings meet, as compared to the first rings with small $\lambda$, resulting in a coarser length scale in the background. Moreover, wetting promotes the continuity of the $\alpha$ domain from an earlier time as opposed to that of neat blends. These factors may have an influence on enhancing the coarsening kinetics of $\alpha$. 
The trends for the asymmetric blends also follow a similar course as shown in Fig.~\ref{4060} where $\alpha$, which is always continuous, shows an enhanced coarsening rate. The effect is more prominent when $\alpha$ also happens to be the majority phase, i.e., $A_{60}B_{40}$. 

\subsection{Coarsening kinetics of the non-wetting phase $\beta$}

\begin{figure}[h]
\centering
\subfloat[]{\includegraphics[scale=0.5]{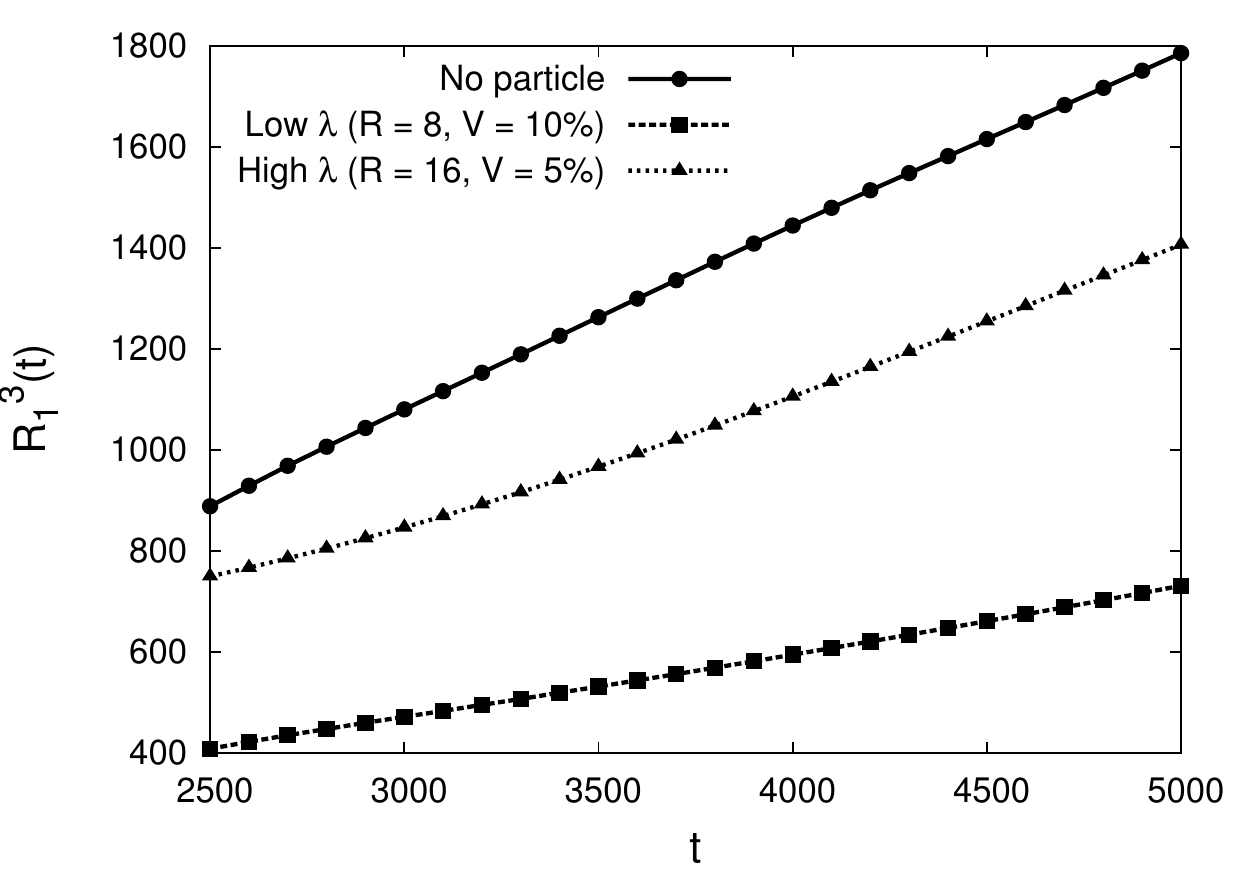}}\hspace{2mm}
\subfloat[]{\includegraphics[scale=0.5]{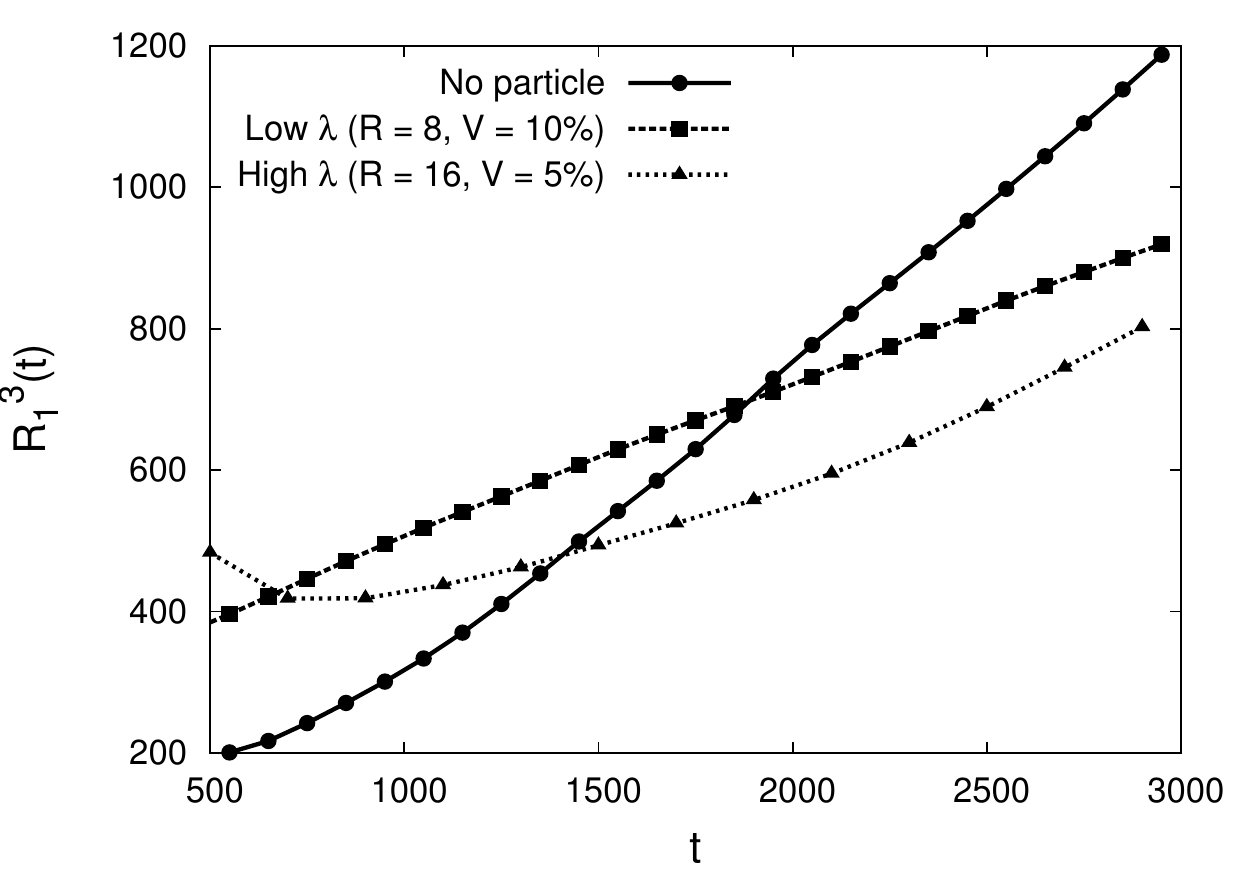}}
\caption{\label{beta_5050} $A_{50}B_{50}$: Average size of the $\beta$ domains $R_1(t)$, given by Eq.~\ref{eq_first_moment}, is plotted against time $t$ in (a) $S_o$ and (b) $S_s$ systems. Note that domain growth follows the LSW $t^{1/3}$ law. Note that timescales are different in the above figures.}
\end{figure}

In a symmetric mixture with neutral interactions, the coarsening kinetics of $\beta$ are similar to that of $\alpha$. As a result, in a spinodally decomposed matrix, the respective domains are of the same size (compare Figs.~\ref{5050}a and \ref{beta_5050}a). In the strongly interacting system $S_s$, on the other hand, coarsening rates of $\beta$ are enhanced at early stages of growth (in Fig.~\ref{beta_5050}b) as it is rapidly expelled from the particle surfaces in order to accommodate $\alpha$. However, with progress in time, $\beta$ no longer remains contiguous and forms isolated chains, and the coarsening rates of $\beta$ are overwhelmed by that of neat blends. Referring to Figs.~\ref{5_16_3000} and \ref{10_8_3000}, note that in large $\lambda$ system, the chains of dispersed $\beta$ phase are thinner when compared to the same with low $\lambda$, where a higher density of particles leads to thicker $\beta$ chains along with thinner continuity of the $\alpha$ matrix. These factors are reflected in the coarsening rates of $\beta$ in Fig.~\ref{beta_5050}b.

\begin{figure}[h]
\centering
\subfloat[]{\includegraphics[scale=0.5]{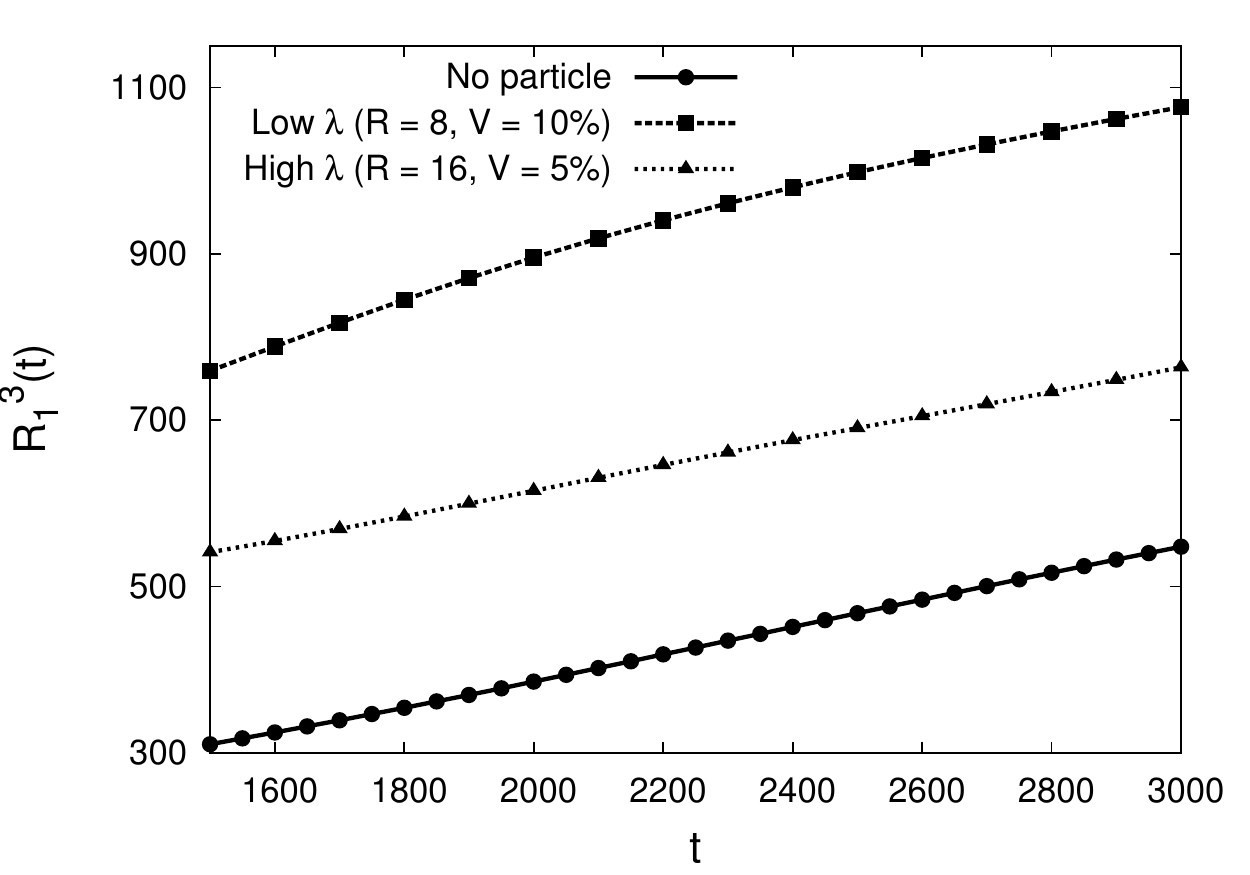}}\hspace{2mm}
\subfloat[]{\includegraphics[scale=0.5]{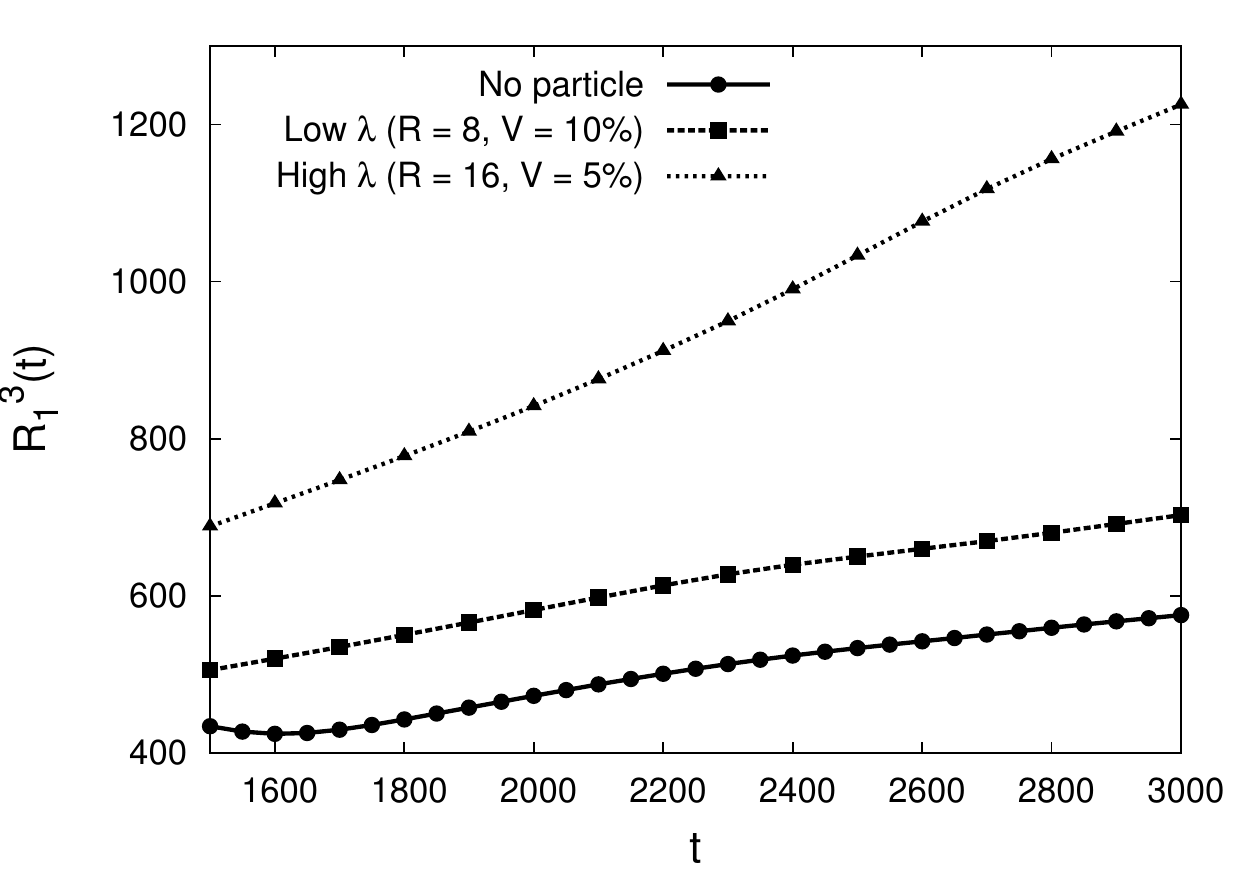}}
\caption{\label{beta_4060} Average size of the $\beta$ domains $R_1(t)$, given by Eq.~\ref{eq_first_moment}, is plotted against time $t$ in (a) $A_{40}B_{60}$ and (b) $A_{60}B_{40}$ $S_s$ systems. Note that domain growth follows the LSW $t^{1/3}$ law.}
\end{figure}

In asymmetric mixtures, $\beta$ exhibits enhanced coarsening rates as compared to that of neat blends in both low and high $\lambda$ systems (see Fig.~\ref{beta_4060}). With low $\lambda$ conditions in $A_{40}B_{60}$ mixture (in Fig.~\ref{o1_10_8_3000}), majority $\beta$ forms a thicker continuous network along with continuous yet thinner minority $\alpha$. $\alpha$ domains, in this particular case, accommodate the $\gamma$ particles inside it, thereby, limiting the coarsening rates of $\alpha$ in a way similar to particle pinning, leading the coarsening rates of $\alpha$ that are lower than $\beta$ (compare Figs.~\ref{4060}a and \ref{beta_4060}a). In the large $\lambda$ system (in Fig.~\ref{o1_5_16_3000}), however, sizes of the $\alpha$ domains are not large enough to accommodate the particles, leading to entrapment of the $\gamma$ particles within the surrounding $\beta$ within an $\alpha$ network; such arrangement of $\beta$ leads to lower coarsening rates than that of $\alpha$, as evident in Figs.~\ref{4060}a and \ref{beta_4060}a.  

In $A_{60}B_{40}$ systems, minority $\beta$ is characterized by isolated droplets in a continuous matrix of $\alpha$ (in Fig.~\ref{o2_ss}). In low $\lambda$ conditions, high density of the particles breaks the $\beta$ phase into many small globular domains, which remain trapped within the interparticle regions, leading to lower coarsening rates (in Fig.~\ref{beta_4060}b) as opposed to that of in large $\lambda$ conditions, where $\beta$ domains coarsen relatively unhindered partially engulfing the particles.

\section{Discussion}\label{discussion}
We begin this section by highlighting four key conclusions from this work:

(a) Importance of interparticle spacing: In strongly interacting system $S_s$, the number of
rings formed around particles is smaller when $\lambda$ is
smaller. Thus, in Fig. \ref{5_16_500}, we find two or three rings
around each particle, while in Fig. \ref{10_8_500}, we find just one or two. 
This causes the bicontinuity to be broken, even in this
symmetric blend, early in the process of microstructure 
formation. While the role of interparticle
spacing is implicit in previous studies, its importance
is revealed in our study quite sharply.
 
(b) Bicontinuity may emerge from a non-bicontinuous microstructures, 
and in asymmetric blends. Even though the early microstructure in an  $A_{40}B_{60}$ blend (in Fig. \ref{o1_10_8_500})
has isolated $\alpha$ filaments (with $\gamma$ particles inside them) embedded inside a continuous background of $\beta$ phase, $\alpha$ filaments connect up by pinching off the $\beta$ phase at its thinnest regions (Fig. \ref{o1_10_8_3000}).

(c) Importance of curvature effects: While the early microstructures
in system $S_{s}$ mimic those expected from surface-directed
SD, curvature leads to the shrinkage and eventual disappearance
of the inner-most ring of $\alpha$ phase, thereby bringing the
(non-preferred) $\beta$ phase in contact with the particle.

(d) Importance of wetting effects: Due to attractive interaction, $\alpha$ wetting layers are formed rapidly (or $\beta$ layers are expelled rapidly) on the particles which further speed up the dynamics of the $\alpha$ (or $\beta$) domains in the spinodally decomposed matrix, resulting in more prominent coarsening rates than that of neat blends.

The ring (or target) pattern we have seen in Figs.~\ref{5_16_500} and \ref{10_8_500}
has been observed in simulations by Lee \emph{et al.}~\cite{Lee}; 
there are at least two key differences between the their study and ours. First, Lee et al 
used a Cahn-Hilliard-Cook model of a binary alloy and added a local surface interaction energy 
for inducing preferential segregation. Thus, while the A-B interfaces are treated in a phase field formalism, 
the particle and the matrix phases share a sharp, Gibbsian interface. Our study, in contrast, uses a Cahn-Hilliard model in a ternary setting which allows us to treat all the three interfaces within the same 
phase field formalism (which has the advantage that it could be extended easily to studying the role of particles with irregular or ramified shapes). Second, Lee et al studied the effect due to a single particle in a symmetric mixture; the present study examines multi-particle effects (in terms of particle size and volume fraction) in both symmetric and asymmetric mixtures.  

The picture that emerges from our results is consistent with previous studies. Target patterns in symmetric mixtures lead to a normal bicontinuous SD pattern at intermediate times. Asymmetric mixtures become bicontinuous or droplet depending on whether the majority or the minority component is preferred to the particles. The microstructures in Figs.~\ref{o1_10_8_500} and ~\ref{o1_10_8_3000} for a $A_{40}B_{60}$ blend are particularly interesting in that particles are localized inside the preferred phase. Such clustered arrangements of particles in the minority phase in a copolymer-particle mixture can be used to design composite architectures~\cite{Thompson2001}. The experiments by Tanaka \emph{et al.}~\cite{Tanaka} and Jiang \emph{et al.}~\cite{Jiang2016} and simulations by Ma~\cite{Ma,Ma2} and Ginzburg \emph{et al.}~\cite{VVBalazs} also produced such clustered microstructures; however, the particle size in their work was far smaller than the length scale of spinodal decomposition. Our simulations indicate that the bulk domains in critical and off-critical mixtures undergo diffusive coarsening as $\sim t^{1/3}$, which is seen in most of the experiments and simulations.

Finally, in the present work we do not consider the hydrodynamic interactions, interparticle interactions,  
or processing conditions (i.e., shear) on the morphological evolution, all of which have a significant influence on phase separation dynamics~\cite{ACBalazs,FCBalazs,DCBalazs}. In addition, we restricted our simulations 
for spherical particles, though our phase field method is not restricted to such geometrically simple shapes. 
Our current simulations, with their emphasis on early and intermediate stages of microstructure 
formation, do not allow us to examine particle effects on late stage evolution due to 
computational limitations. Our ongoing work focuses on the influence of various particle configurations and particle aggregates such as fractal surfaces on the formation and stability of interference patterns due to interaction of the composition waves about the particles; by engineering the locations and structures of the particle phases, phase separating morphology of the mixtures can be designed and controlled~\cite{Jiang}.  

\section{Summary and Conclusions}\label{summary}
\begin{enumerate}
\item We have studied the effect of immobile particles on phase separation microstructures in ternary mixtures through computer simulations based on a ternary phase-field model in which the particles which are C-rich are embedded in an A-B blend.
\item We have explored a regime of interparticle separation distances  (a few times the characteristic length scale of spinodal decomposition) in which interesting new effects may be expected.
\item Our study shows four new effects in systems in which the particle phase has a strong preference for one of the components. 
\item Microstructures in a symmetric blends are not bicontinuous in the presence of strongly interacting particles.
\item Initially non-bicontinuous microstructures in asymmetric blends may evolve to become bicontinuous at intermediate times.
\item Even though the particle phase may be wetted by the preferred component, this wetting layer may dissolve due to curvature-driven coarsening leaving the particle in contact with the non-preferred component.
\item $\alpha$ and $\beta$ domains in the bulk patterns coarsen at intermediate times and scale as $R_1(t) \sim t^{1/3}$. While continuous wetting phase $\alpha$ exhibits enhanced coarsening rates due to its preference to the $\gamma$ particles, dispersed non-wetting phase $\beta$ within the interparticle regions exhibits lower coarsening rates.
\end{enumerate}


\section*{Appendix: Calculation of Interfacial Energies}\label{appendix}
Interfacial energy between phases (say, $\alpha$-$\beta$) can be calculated assuming a flat interface and following Eq. \ref{2} as
\begin{equation}\label{a1}
\sigma_{\alpha\beta} = N_v \int_{-\infty}^{\infty}\left[\Delta f(c) + \textstyle \sum_i \kappa_{i}\left(\nabla c_i\right)^2 \right]\, dx ,
\end{equation}  
where $x$ is a position perpendicular to the interface and
\begin{equation}\label{a2}
\Delta f(c) = f(c_i) - \textstyle \sum_i \, c_i \, \mu_i^{\alpha/\beta}, \hspace{0.5cm} i = A, B, C.
\end{equation}
$c_i$ are the equilibrium solutions of Eqs. \ref{5} and \ref{6}. $\mu_i^{\alpha/\beta}$ is the chemical potential of component $i$ in $\alpha$ and $\beta$ phases in chemical equilibrium (chemical potential matching for each component), which can be derived from suitable derivatives \cite{Lupis, Ghosh} of Eq. \ref{3}. $\sigma_{\beta\gamma}$ and $\sigma_{\alpha\gamma}$ can also be evaluated in a similar way. For more details, please refer to \cite{Ghosh,Huang1999}.
\bibliographystyle{unsrt}
\bibliography{spinodal}
\end{document}